\RequirePackage{rotating}

\documentclass[useAMS,usenatbib]{mn2e}

\usepackage{graphicx}

\title[Spectropolarimetry of nuclear winds]{Spectropolarimetry of NGC~3783 and Mrk~509:\\
Evidence for powerful nuclear winds in Seyfert 1 Galaxies}

\author[P. Lira et al.]{P. Lira$^{1}$, M. Kishimoto$^{2}$, R. W. Goosmann$^{3}$, R. Campos$^{1}$, D. Axon\thanks{Deceased on the 5th April 2012},
  M. Elvis$^{4}$,  \newauthor A. Lawrence$^{5}$, B. M. Peterson$^{6,7}$, A. Robinson$^{8}$\\
$^{1}$Departamento Astronom\'ia, Universidad de Chile, Casilla 36D, Santiago, Chile\\
$^{2}$Department of Astrophysics \& Atmospheric Sciences, Kyoto Sangyo University, Kyoto 603-8555, Japan\\
$^{3}$Observatoire Astronomique de Strasbourg, Université de Strasbourg, CNRS, UMR 7550, 11 rue de l'Université, F-67000 Strasbourg, France\\
$^{4}$Harvard-Smithsonian Center for Astrophysics, 60 Garden St., Cambridge, Massachusetts 02138 USA.\\
$^{5}$Institute for Astronomy, SUPA (Scottish Universities Physics Alliance), University of Edinburgh, Royal Observatory, Blackford Hill, Edinburgh EH9 3HJ, UK\\
$^{6}$Department of Astronomy, The Ohio State University, 140 West 18th Avenue, Columbus, OH 43210, USA\\
$^{7}$Center for Cosmology and AstroParticle Physics, The Ohio State University, 191 West Woodruff Ave., Columbus, OH 43210, USA\\
$^{8}$Laboratory for Multiwavelength Astrophysics and School of Physics and Astronomy, Rochester Institute of Technology, Rochester 14623, New York, USA}

\begin{document}

\date{}

\pagerange{\pageref{firstpage}--\pageref{lastpage}} \pubyear{2002}

\maketitle

\label{firstpage}

\begin{abstract}

We present results from high signal-to-noise optical
spectropolarimetric observations of the Seyfert 1 galaxies NGC\,3783
and Mrk\,509 in the 3500-7000 \AA\ range. We find complex structure in
the polarized emission for both objects. In particular, Position Angle
(PA) changes across the Balmer lines show a distinctive {\it
  'M'-shaped\/} profile that had not been observed in this detail
before, but could represent a common trait in Seyfert 1 galaxies. In
fact, while this shape is observed in all Balmer lines in NGC\,3783,
Mrk\,509 transitions into a {\it 'M'-shaped\/} PA profile for higher
transitions lines. We have modeled the observed profiles using the
STOKES radiative transfer code and assuming that the scattering region
is co-spatial with the BLR and outflowing. The results give compelling
new evidence for the presence of nuclear winds in these two Seyfert 1
galaxies.
 
\end{abstract}


\section{Introduction}

Arguably, the results from spectropolarimetry of Active Galactic
Nuclei (AGN) has been the primary observational evidence that has led
to the acceptance of the standard `Unified Model' of AGNs: to a first
order, the accretion disk is surrounded by a similarly oriented
optically thick dusty torus, with the collimated radio emission being
coincident with the system axis. The Broad Line Region (BLR) would
also lie within the opaque torus and for Type~1 AGN a direct view of
the nuclear source and BLR is only possible in the polar direction.
For Type~2 AGNs the BLR can be observed in polarized light, after
scattering takes place somewhere in the vicinity of the central
source. The position angle (PA) of this polarized emission is
generally perpendicular to the radio axis of the source. The rather
flat wavelength dependence of the continuum polarization suggests that
the nuclear light is scattered by electrons, while the observed PAs
indicate that the scattering occurs somewhere along the throat of a
torus, i.e., in the polar directions as seen from the nucleus.

Even Type~1 AGNs sometimes show the same evidence for such 'polar
scattering' (Smith et al., 2002). However, spectropolarimetric studies
of Type~1 AGNs have found that a significant fraction presents
continuum position angles which are, to first order, parallel rather
than perpendicular to the radio axis of the system, suggesting that
besides the 'polar scatterer' there is another, much more compact
'equatorial scattering' region, that can dominate the polarized
spectrum in a significant fraction of Seyfert 1 galaxies (Smith et
al., 2002, 2004).
 
But, more importantly, the spectropolarimetric results show that in
sources with an equatorial scattering region, the polarization PA
often rotates significantly across the broad emission lines. This
happens for near-field scattering, i.e., with the scatterer being
close enough to the H$\alpha$ emitting region to resolve it so that
the radiation from the two opposite sides of the, presumably flattened
BLR, is scattered at different angles. As the level of polarization in
Type 1 AGN is rather low ($\sim 1\%$), previous studies using 4-meter
class telescopes have had a limited number of sources accessible to
spectropolarimetry of adequate signal-to-noise. This problem is
particularly acute in the blue region of the optical spectrum.

We have selected two bright Seyfert 1 galaxies with a known degree of
polarization to further study the characteristics of their polarized
emission using 8-meter facilities. NGC\,3783 and Mrk\,509 are two well
studied, southern Seyfert 1 galaxies for which emission-line lags for
several BLR lines have been measured from reverberation mapping
(Reichert et al. 1994; Stirpe et al. 1994; Onken \& Peterson 2002,
Peterson et al.~2004). Updated black hole masses for both Seyferts can
be found in the AGN Black Hole Mass Database (Bentz \& Katz,
2015). Hence, {\it physical\/} scales for various line emitting
regions are available. This is potentially of great importance in
understanding the near-field polarization since it could lead to a
determination of a physical size for the scatterer and, hopefully, the
identification of its nature. Spectropolarimetry of Mrk\,509 has
already been presented by Goodrich \& Miller (1994), Young et al.\,
(1999), Schmid et al.\, (2000), Smith et al.\, (2002), and Afanasiev
et al.\,(2019), while previous results for NGC\,3783 have been
presented by Smith et al.\, (2002).

\section{Observations and Data Reduction}

\subsection{NGC\,3783 observations}
 
We observed NGC\,3783 (z=0.009730) on the 3rd of April 2006 using the
FORS1/600B spectropolarimetry mode on the Unit Telescope 2 of the VLT
(resolution $\sim 6$ \AA), covering the H$\beta$ and higher order
Balmer emission lines, and on the 30th of April 1st of May 2006 using
EFOSC2/Gr\#4 on the 3.6m telescope at La Silla (resolution $\sim 12$
\AA), covering the H$\alpha$ emission line.\footnote{Details of the
  instrument optical components can be found here
  https://www.eso.org/sci/facilities/paranal/instruments/fors/inst/pola.html
  and here
  https://www.eso.org/sci/facilities/lasilla/instruments/efosc/inst/Efosc2PolarElements.html}

We implemented sequences of observations consisting of four frames
with different waveplate positions (0, 45, 22.5 and 67.5 degrees). The
total exposure time was 3.4 hours with FORS 1 and 6.2 hours using
EFOSC2. At the VLT, the 1.5\arcsec\ slit was fixed in the north-south
direction, and the atmospheric dispersion compensator was reset at the
start of each sequence. At the 3.6m the slit was approximately
parallel to the parallactic angle. No observations were taken at
airmass in excess of 1.2. 

The CCD frames were bias-subtracted using an averaged bias frame, and
flat-fielded using the flat field frames taken with external
calibration units during day time. The pixels with obvious cosmic ray
hits were fixed using neighboring pixels. There was no need to correct
for CCD distortions previous to the spectrum subtraction as o- and
e-rays were located close to each other on the detectors ($\sim 116$
pixels $\sim 23'' \sim 2.8$ mm for FORS1 and $\sim 62$ pixels $\sim
10'' \sim 4.1$ mm for EFOSC2). Inspection of sky lines showed that
these had a standard deviation of 0.4 and 0.6 pixels for FORS1 and
EFOSC2, respectively, when comparing positions above and below the
double spectra and for a wide range of airmasses. Sky subtraction was
achieved in the standard way by defining windows on either side of the
object spectra. Notice that the individual slits for the o- and e-rays
were wide enough to be able to determine the sky level from the same
slit where the spectra were found. No residuals were observed at the
positions of the sky lines. The sky-subtracted o- and e-ray spectra
were combined to produce normalized Stokes parameters $q$ and $u$ and
the Stokes parameter $I$, following Miller, Robinson \& Goodrich
(1988). The polarization level and position angle, $p$ and PA, were
obtained in the standard way ($p = \sqrt(q^2+y^2)$, PA $=
1/2\ \tan^{-1}(u/q)$). $p$ was also obtained using 'unbiased'
prescriptions (Miller, Robinson \& Goodrich 1988), but the resulting
spectra did not show any significant differences, so the standard
definition was finally adopted. The zero point of the polarization
signal was checked using the polarized standard star Hiltner\,652
($p=6.3\%$ at 5500 \AA) in both runs. The measurement of unpolarized
standard stars showed a low instrumental polarization: in the
5200-5800 \AA\ range HD94851 gave $p=0.07\pm0.02$ (\%) for FORS1
(VLT), while HD97689 gave $p=0.02\pm0.01$ (\%) for EFOSC2 (3.6m).

No order-sorting filter was available for the EFOSC2 observations, and
thus the measurements at long wavelengths might be slightly affected
by a second-order contamination. However we note that the observation
of the blue unpolarized star above did not show any systematic
polarization at $\lambda \la 7200$ \AA.

Flux calibration was achieved using the flux standard stars EG274
(VLT), and LTT4816 and Feige67 (3.6m). The data were corrected for
Galactic extinction assuming $E_{B-V} = 0.119$ and using the
extinction curve derived by Cardelli, Clayton, \& Mathis (1989) with
$A_{V}/ E_{B-V} = 3.1$.

 \begin{figure*}
 \centering
 \includegraphics[scale=1.2,angle=0,trim=50 0 0 0]{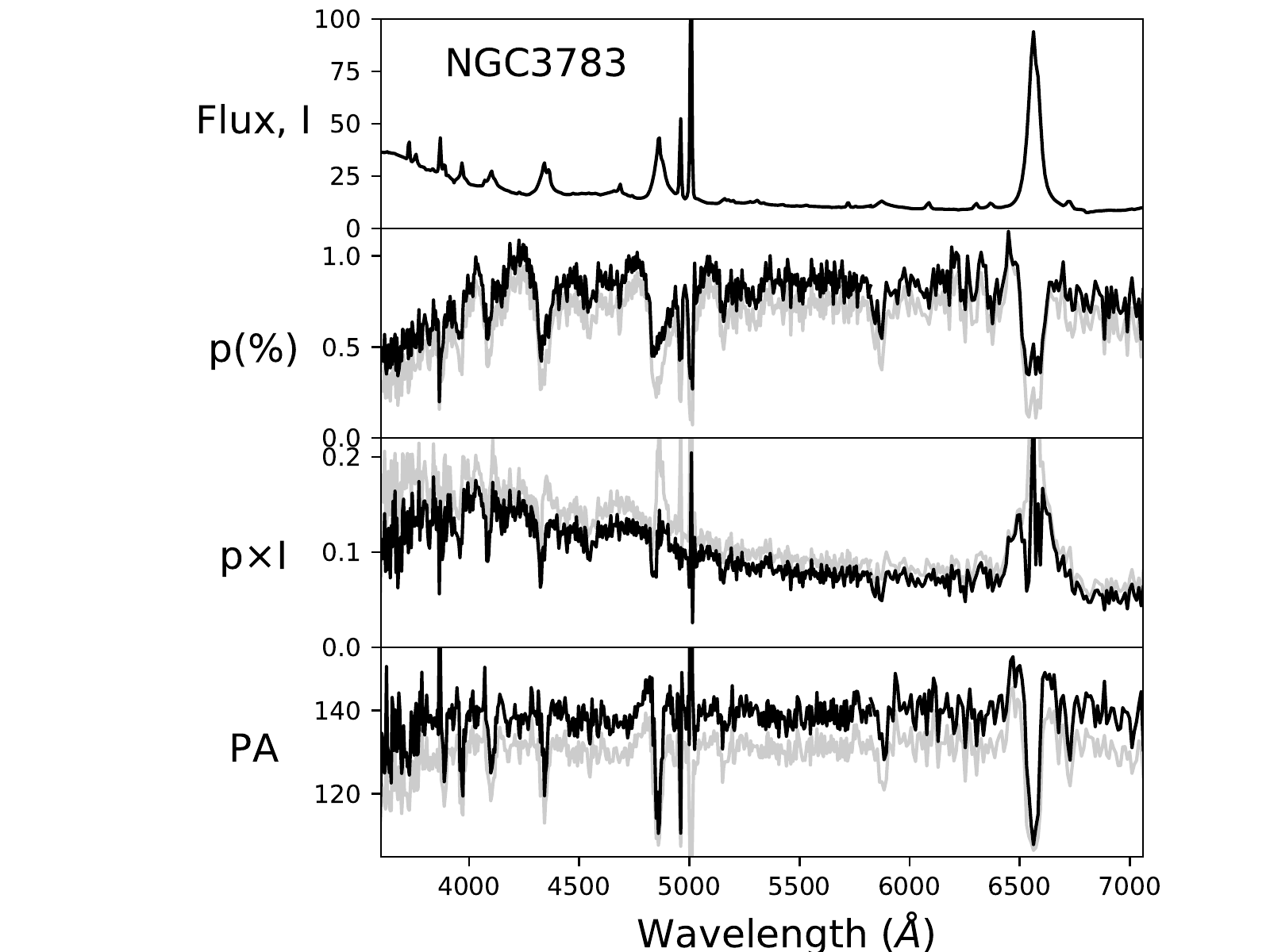}
 \caption{Rest-frame spectropolarimetry of NGC\,3783. From top to
   bottom: total flux ($I$ -- in units of $10^{15}$ erg/s/cm$^2$/\AA),
   degree of polarization ($p$ -- in percentage), polarized flux
   ($p\times I$ -- in units of $10^{15}$ erg/s/cm$^2$/\AA) and
   polarization position angle (PA -- in degrees). Black spectra
   correspond to data corrected for intervening ISP polarization,
   while gray spectra correspond to the data as observed.}
 \end{figure*}

 \begin{figure*}
 \centering
 \includegraphics[scale=1.2,angle=0,trim=50 0 0 0]{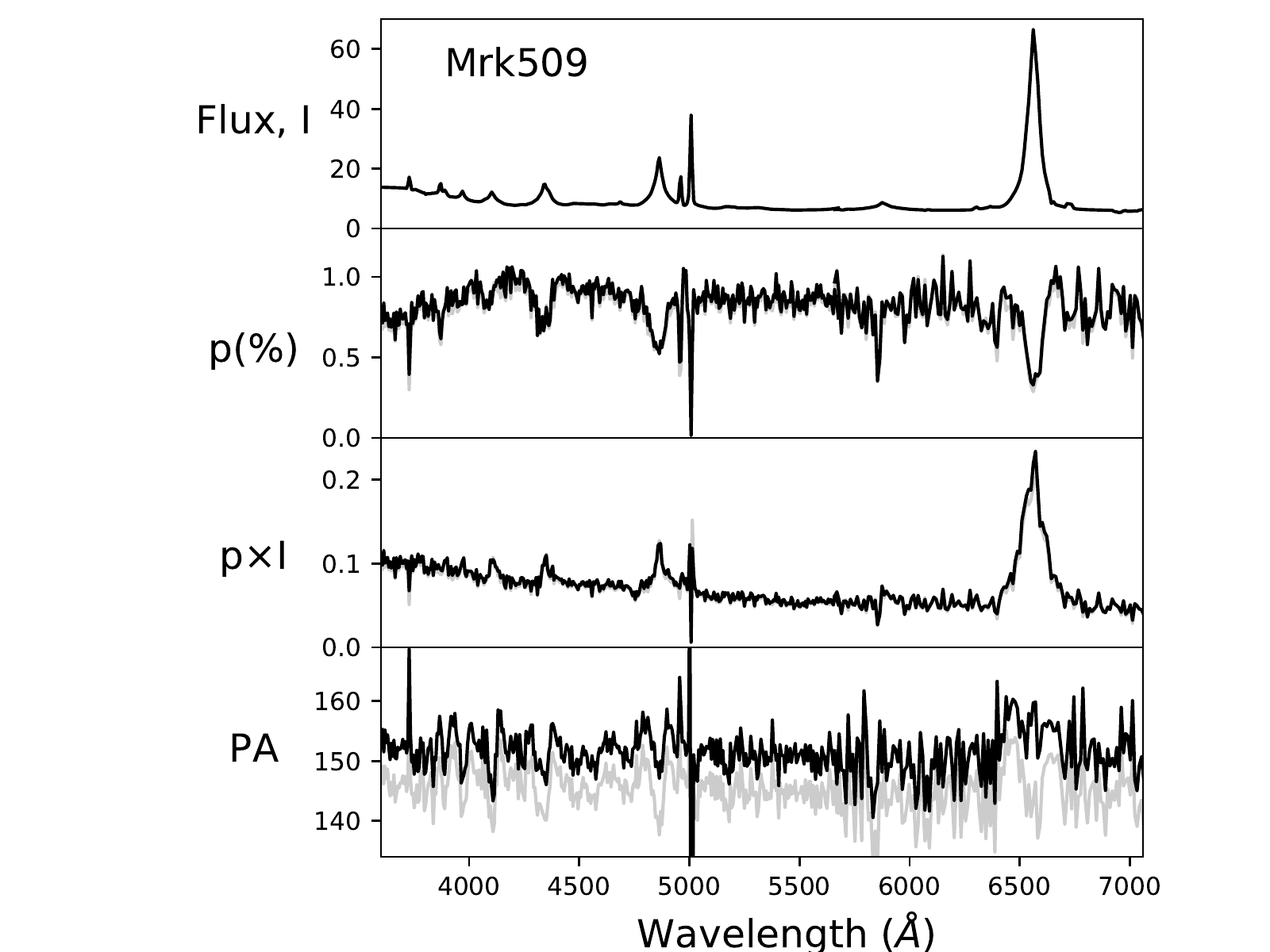}
 \caption{Rest-frame spectropolarimetry of Mrk\,509. From top to
   bottom: total flux ($I$ -- in units of $10^{15}$ erg/s/cm$^2$/\AA),
   degree of polarization ($p$ -- in percentage), polarized flux
   ($p\times I$ -- in units of $10^{15}$ erg/s/cm$^2$/\AA) and
   polarization position angle (PA -- in degrees). Black spectra
   correspond to data corrected for intervening ISP polarization,
   while gray spectra corresponds to the data as observed.}
 \end{figure*}
 
\subsection{Mrk\,509 observations}

We observed Mrk\,509 (z=0.034397) on the 23th of August 2006 using
FORS1 on the UT2 at the VLT. Exposure times were 6.2 and 0.3 hours,
for the 600B and 600R grisms (resolution $\sim 6$ \AA\ for both
settings), respectively. No observations were taken at airmass in
excess of 1.5.

Images were reduced in the same manner as described above for
NGC\,3783. The zero point of the polarization signal was checked using
the polarized standard star BD-125133 ($p=4.4\%$ at 5500 \AA). The
star EGRR 274 was used as unpolarized and flux standard. The
measurement confirmed the low instrumental polarization at
$p=0.10\pm0.05$ (\%) in the 5200-5800 \AA\ range. As before, the data
were also corrected for a foreground extinction of $E_{B-V}=0.057$.

\subsection{Intervening Polarization and Atmospheric Depolarization}

An important concern when analyzing polarimetric data is that
dust-induced Inter-Stellar Polarization (ISP) along the line of sight
to the active nucleus must be taken into account (Hoffman et al.,
2005).

To quantify the amount of ISP in our Galaxy towards NGC\,3783
($b=23$\degr), we observed a few Galactic stars in the same
field. They are close in projection to NGC\,3783 and have
spectroscopically-determined distances large enough to probe most of
the interstellar polarization within the Galactic disk. These
observations suggest the Galactic interstellar polarization towards
NGC\,3783 to be $\sim$ 0.25\% at PA $\sim$105\degr. The same exercise
along the line of sight towards Mrk\,509 ($b = -29$\degr) gives a
Galactic interstellar polarization of $\sim$0.18\% at PA $\sim$
100\degr.

We examined the level of polarization in the OIII lines after ISP
correction. For this, it is necessary to first subtract the continuum
in the original polarized flux spectra, which was achieved by fitting
a low degree polinomium to the data. The results show that the
intrinsic polarization of these lines is very low, at $p \sim 0.001
\pm 0.0003$\% for NGC\,3783 and $p \sim 0.002 \pm 0.0006$\% for
Mrk\,509, in agreement with their noisy appearance in the $p \times I$
spectra of both sources. This is expected considering that the Narrow
Line Region (NLR) is sufficiently large so that geometrical
cancellation (this is, the depolarization due to the combination of
rather random polarized signals) can be very efficient. This is a good
indication that there are no further sources of intervening ISP
contamination towards the nucleus of these two galaxies. Corrected
and uncorrected observations are presented in Figures 1 and 2.

Finally, we also tried to correct for the amount of atmospheric
depolarization affecting our observations. As explained by Afanasiev
\& Amirkhanyan (2012), the atmosphere introduces a variable amount of
transmission between consecutive frames, which leads to a
time-dependent differential intensity in the o- and e-ray spectra
obtained from different frames. As shown by Afanasiev \& Amirkhanyan
(2012) the time scale for these changes is about 30 min and appears as
low amplitude oscillations in the Stokes spectra. As repeat visits to
our sources take about the same time scale (individual exposures were
400 seconds for NGC\,3783 and 500 seconds for Mrk\,509, but it is
required to cycle through four frames with different waveplate
positions in order to determine the Stokes vectors), it was not
feasible to correct for this effect. However, as we will show in the
next section, the amplitude of the features observed in the spectra of
NGC\,3783 and Mrk\,509 are not only above those shown by Afanasiev \&
Amirkhanyan (2012) due to atmospheric depolarization, but also they
strictly coincide with the presence of emission lines in the spectra
and have common patterns, therefore demonstrating that they must be
intrinsic in nature.

\section{Results}
 
Figures 1 and 2 show the total flux ($I$), degree of polarization ($p$
-- in percentage), polarized flux ($p\times I$) and polarization
position angle (PA) obtained from our observations of NGC\,3783 and
Mrk\,509, before and after applying the corrections for polarization
along the line of sight detailed in Section 2.3. In what follows we
will discuss the main results from the ISP corrected observations.

\subsection{Common Features}

The median value of the corrected PA spectra gives important
information about the location of the scatterer medium when it can be
contrasted with information about the geometry of the systems. In
Figure 3 we show the mean PA obtained through our spectropolarimetric
observations and the projected axis of symmetry of the systems as
obtained from dynamical evidence for NGC3783 (Fischer et al.~2013) and
radio observations for Mrk509 (Singh \& Veestergard, 1992). In both
cases we see that the angles are almost aligned, confirming that these
two objects are dominated by equatorial scattering.

Our observations of both Seyfert 1 galaxies also show that the degree
of continuum polarization $p$ is essentially flat above $\sim
4000$ \AA\, arguing that electron scattering is responsible for the
observed polarization. The $p$ spectra also show strong decrements at
the position of the core of the emission lines, indicating that the
lines have a lower fraction of polarized flux than the
continuum. There is also evidence, particularly clear in the case of
NGC\,3783, that the broad Balmer lines are flanked by high
polarization shoulders coincident with the line wings. This is similar
to what is found in other Seyfert galaxies, where the broad line
components seem more polarized than the surrounding continuum (e.g.,
Schmid et al., 2000), resulting in broader emission lines in polarized
flux, as has been already discussed by Goodrich \& Miller (1994) and
Young et al.\ (1999) for the case of Mrk\,509.

As observed by Young et al.\ (1999) the lower polarization at the line
cores is indicative of an intrinsic lower polarization or of
polarization at a significantly different position angle. To explain
this further, consider the intensity of the polarized continuum and
line flux as $f_{\rm cont}$ and $f_{\rm line}$, respectively. Then for
a $\Delta {\rm PA} \sim 0\degr$, $p\times I \sim f_{\rm cont} + f_{\rm
  line}$; but if $\Delta {\rm PA} \sim \pm 90\degr$, then $p\times I
\sim \pm (f_{\rm cont} - f_{\rm line})$.  For $0 < \Delta {\rm PA} <
\pm 90\degr$, intermediate cases would be found\footnote{Note that
  this happens because the two signals are incoherent, and therefore
  their sum is {\em not} the result of the vectorial addition of plain
  waves, but of their Stokes vectors.}. Hence, depolarization occurs
for $\Delta {\rm PA} \neq 0$, and therefore it is a direct consequence
of a different spatial distribution of the emitting and/or scattering
regions\footnote{In the modelling presented in Section 4,
  depolarization will result from the different spatial distribution
  of the continuum and line emitting regions.}. Our observations
clearly show that depolarization strongly affects the broad components
of the emission lines.

PA rotation across the broad lines has long being recognized as common
in Seyfert 1 galaxies (Miller \& Goodrich, 1994; Young et al., 1999;
Schmid et al., 2000; Smith et al., 2002, 2004) and Afanasiev et
al.\,(2019) recently published a compilation of spectropolatimetric
data for 30 Seyfert 1 galaxies showing H$\alpha$ PA rotation from
which black hole masses were derived. We observe PA rotation in both,
Mrk\,509 and NGC\,3783 and will discuss the details in Sections 3.2
and 3.3.

The lower polarization below 4000 \AA\ seen in Figures 1 and 2 is in
striking contrast with the behavior of the continuum observed in
direct light, which rises steeply towards the blue in both
galaxies. This has not been noticed in other Seyfert 1 galaxies
before, although a marginal slope change in $p$ might be present in
previous observations of objects with spectra obtained below 4000 \AA:
Mrk\,376, Mrk\,704, I\,Zw\,1 (Goodrich \& Miller 1994; Smith et al.,
1997), the exception being Fairall\,51, whose $p$ spectrum raises
continuously towards the blue (Schmid et al., 2001; Smith et al.,
2004).

A similar behavior to that observed in the $p$ spectra of NGC\,3783
and Mrk\,509 has already been observed in quasars (Antonucci 1988;
Schmidt \& Smith et al., 2000; Kishimoto, Antonucci \& Blaes, 2003),
and has been suggested as due to the presence of 'small blue
bump' \footnote{The 'small blue bump' corresponds to the Balmer
  continuum and a blend of higher order Balmer and FeII emission lines
  (Wills, Netzer \& Wills 1985).} (SBB) emission, which can dominate
the continuum flux below 4000 \AA. The SBB should be polarized in the
same way as the BLR, as it appears as direct consequence of the
emission line formation processes. Therefore, this pseudo-continuum
will depolarize the continuum in the same manner as the broad emission
lines, resulting in a lower polarized flux at the blue end of the
spectra.

\begin{figure}
\centering
\includegraphics[scale=0.4]{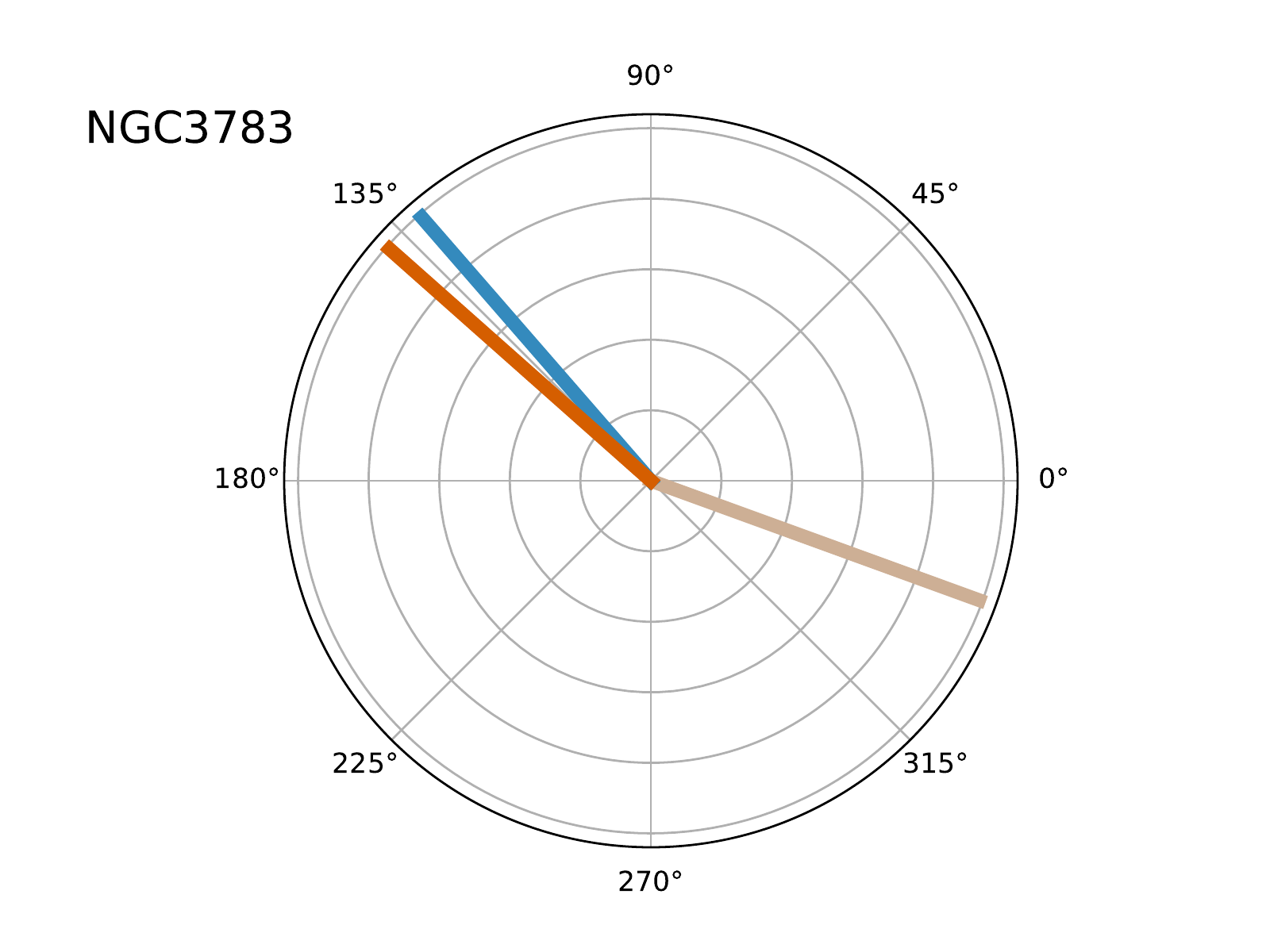}
\includegraphics[scale=0.4]{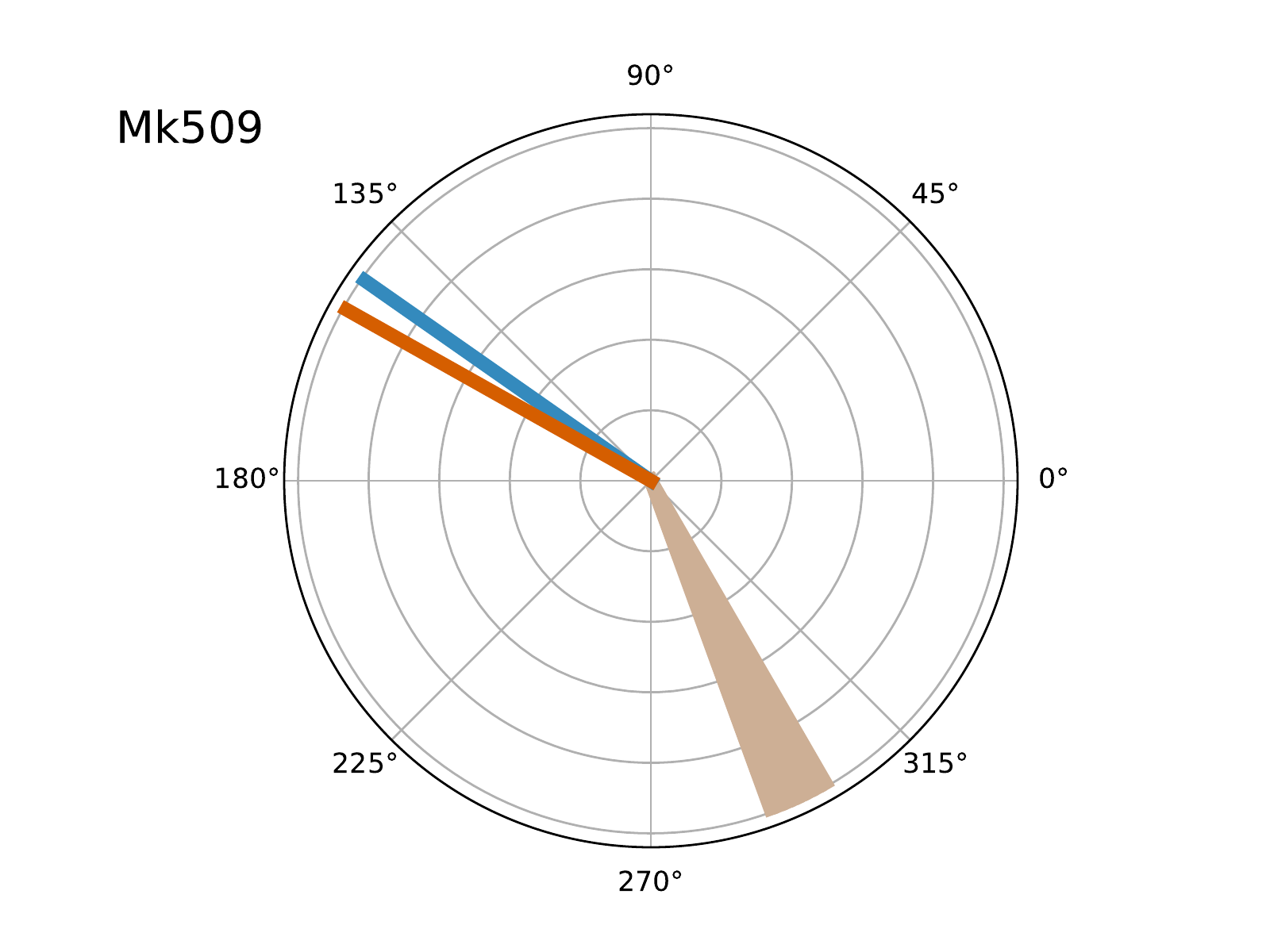}
\caption{Comparison of the mean PA values determined from
  spectropolarimetry and other methods for our sources. The axis of
  symmetry of the systems as determined by Fischer et al.~(2013) for
  NGC\,3783 and Singh \& Veestergard (1992) for Mrk\,509, are shown in
  brown. Position Angles (PAs) derived by our observations are
  presented in blue (uncorrected values) and orange (corrected by
  Galactic ISP).}
\end{figure}

 \begin{figure*}
 \centering
 \includegraphics[scale=0.65,angle=0,trim=0 0 0 0]{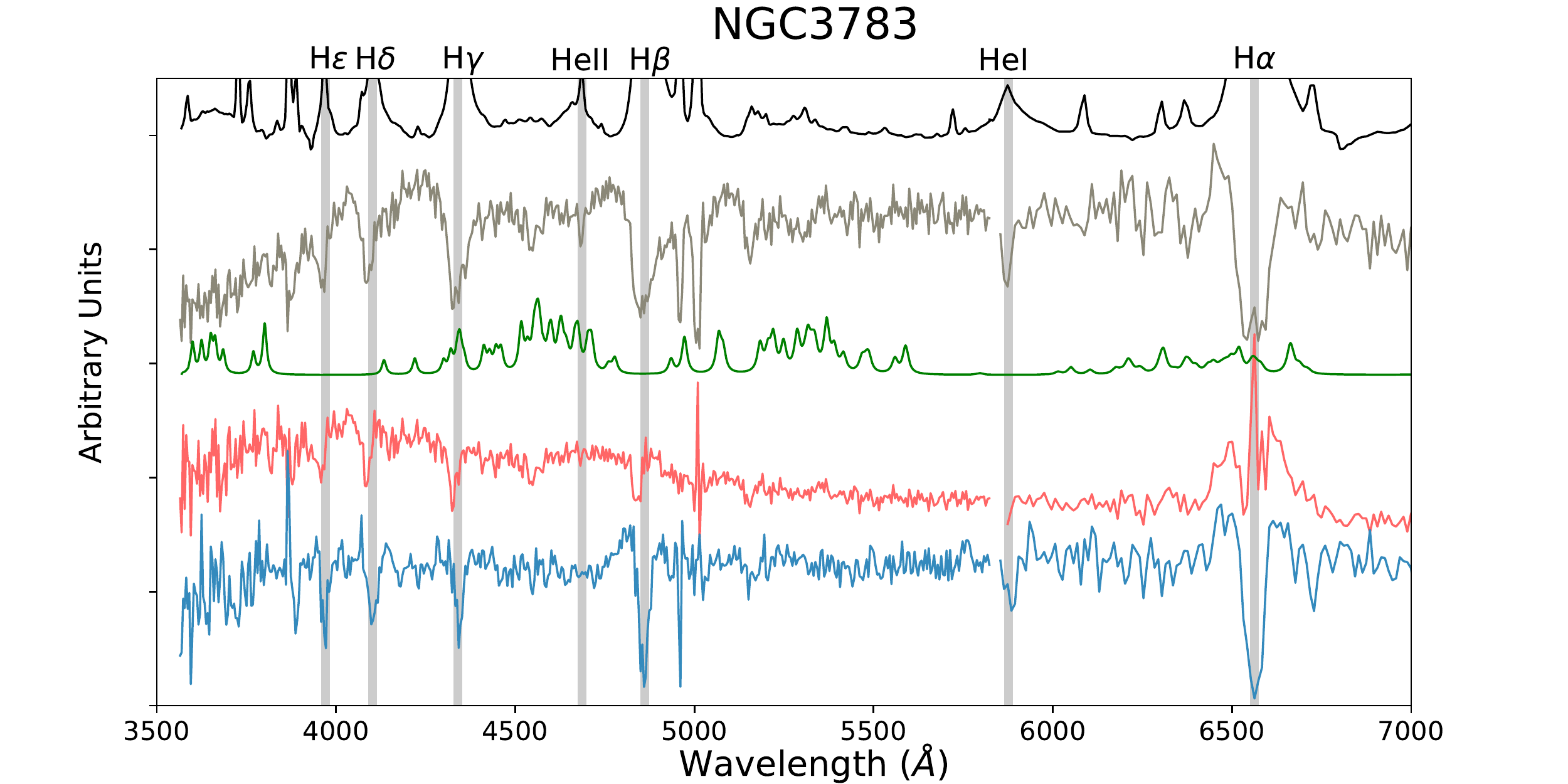}\\
 \includegraphics[scale=0.65,angle=0,trim=0 0 0 -40]{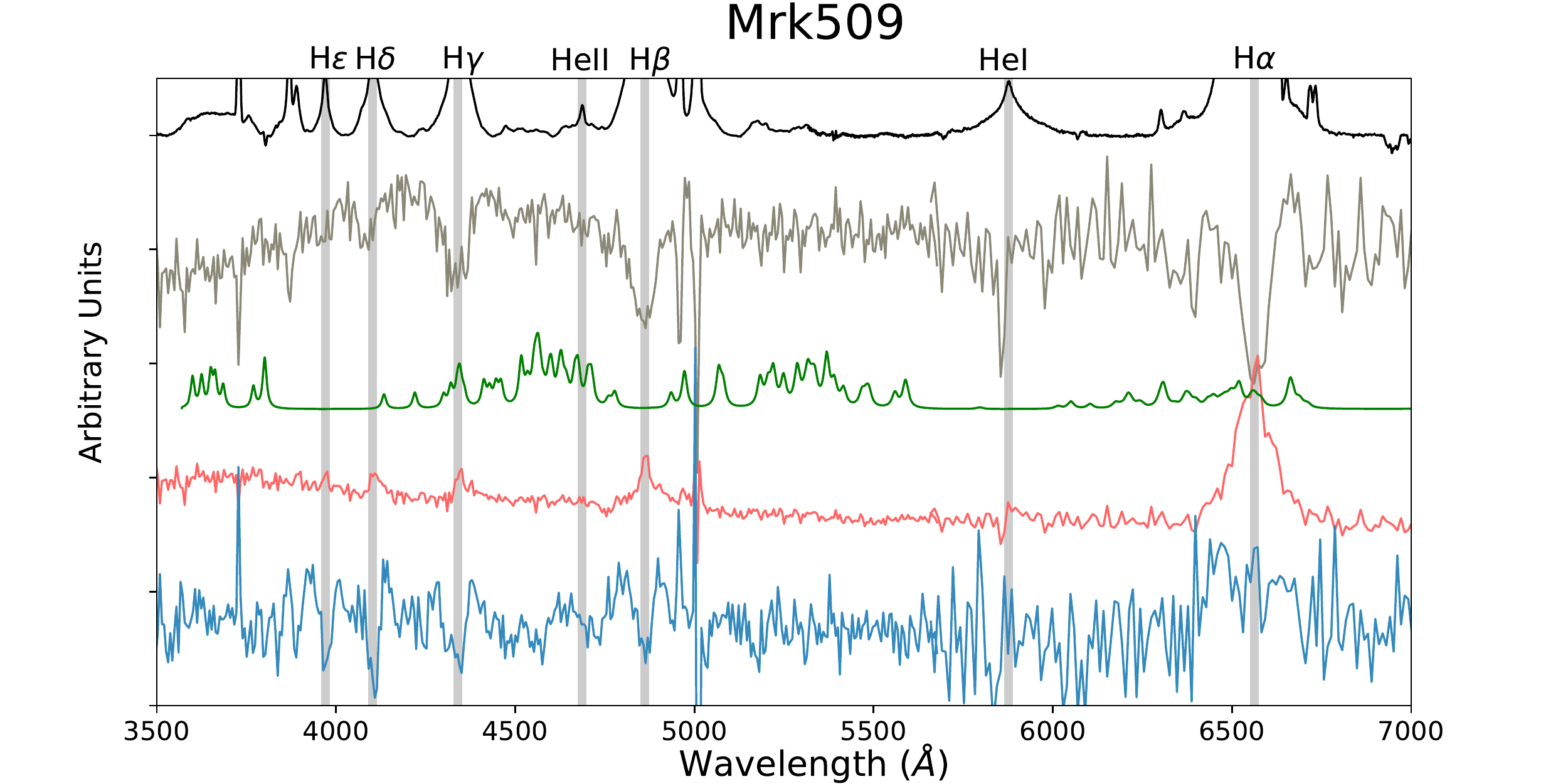}
 \caption{$I$ (black), $p$ (grey), $p\times I$ (red), and PA (blue)
   spectra for NGC\,3783 (top) and Mrk509 (bottom). A FeII template
   has also been included (green). The $I$ spectra have been continuum
   subtracted to allow for a better inspection of the line
   positions. The FeII flux template is based on I Zw 1 (kindly
   provided by M.~Vestergaard; for more details see Vestergaard \&
   Peterson 2005). Vertical lines have been drawn to coincide with the
   central wavelength of the Balmer and He emission lines.}
 \end{figure*}

\subsection{NGC\,3783}

Essentially no variation was found between the data obtained during
our VLT and 3.6m runs, which overlap in the 4060-5870 \AA\
region. Hence, in what follows we directly compare the structure
observed in the H$\alpha$ and H$\beta$ emission lines obtained in the
different runs.

The observed continuum polarization level is clearly higher than that
observed by Smith et al.~(2002), who reported observations obtained
with the AAT. Our data (Apr 2006) show $p \sim$ 0.9\% at $\sim
131\degr$ {\it before\/} ISP correction, while Smith's observations
(May 1997; no ISP correction introduced) show $p \sim 0.5\%$ at $\sim
136\degr$. After correcting for ISP (see Section 2.3), the mean
polarization is found to be $p = 0.7 \pm 0.1\%$ in the 5300-5600
\AA\ range. The PA spectrum is centered around $138 \pm
2.5\degr$. H$\alpha$ and H$\beta$ clearly show a {\it `M-shaped'\/} PA
profile, with very deep central troughs. In higher order Balmer lines,
the `shoulders' of the $M$ feature might however not be present, but
the central deep troughs are still very conspicuous. This peculiar
shape had already been seen in the H$\alpha$ line presented in the
work of Smith et al.~(2002), who interpreted it as possibly due to the
effects introduced by the NII doublet. This is clearly not the case,
as we have shown that the NLR shows negligible polarization and the
same profile is repeatedly seen in all Balmer transitions.

Figure 4 shows the $I$, $p$, $p \times I$ and PA spectra for NGC\,3783
in arbitrary units. The $I$ spectrum was continuum subtracted using a
low-degree spline polynomial fit for better inspection of the base of
the emission line features. A FeII template has been also included
(Vestergaard \& Peterson 2005).

The presence of depolarization features at the location of emission
lines and the SBB is more clearly appreciated in Figure 4. For
example, there seems to be hint of depolarization coincident with the
HeII line at 4686 \AA, while a much stronger signal is seem for HeI at
5876 \AA. A slight decrease in $p$ is also seen at the location of the
FeII emission line blends located on each side of the H$\beta$ line
($\sim$ 4500-4600 \AA, and $\sim$ 5150-5350 \AA), although no clear
change of the position angle appears to be present at these locations.

In the $p\times I$ spectrum the FWHM of H$\alpha$ is very broad, at
$\sim 220$ \AA, almost three times the width of the line seen in total
flux. Together with the coincidence of the PA with the projected axis
of symmetry of the system, this is confirmation of scattering taking
place in the equatorial plane of the system, as the velocity
difference between the BLR and scattering material is not diminished
by projection effects (Smith et al., 2005). The $p\times I$ spectrum
is also fairly unusual in that the Balmer lines are seen only `in
absorption', with the exception of H$\alpha$. Clearly, the
depolarization in NGC\,3783 at the line positions is very strong,
suggesting that the lines and the continuum are polarized at very
different angles. This is consistent with the very strong PA changes
($\sim 30\degr$) seen across the lines.

Notice also how the depolarization of the Balmer lines is blue-shifted
with respect to the peak position of the lines seen in total
flux. This can be seen when comparing the $p$ and $p \times I$ spectra
presented in Figure 4. The PA troughs, however, are symmetric with
respect to line centers.

\subsection{Mrk\,509}

 \begin{figure}
 \centering
 \includegraphics[scale=0.38,angle=0,trim=40 0 0 0]{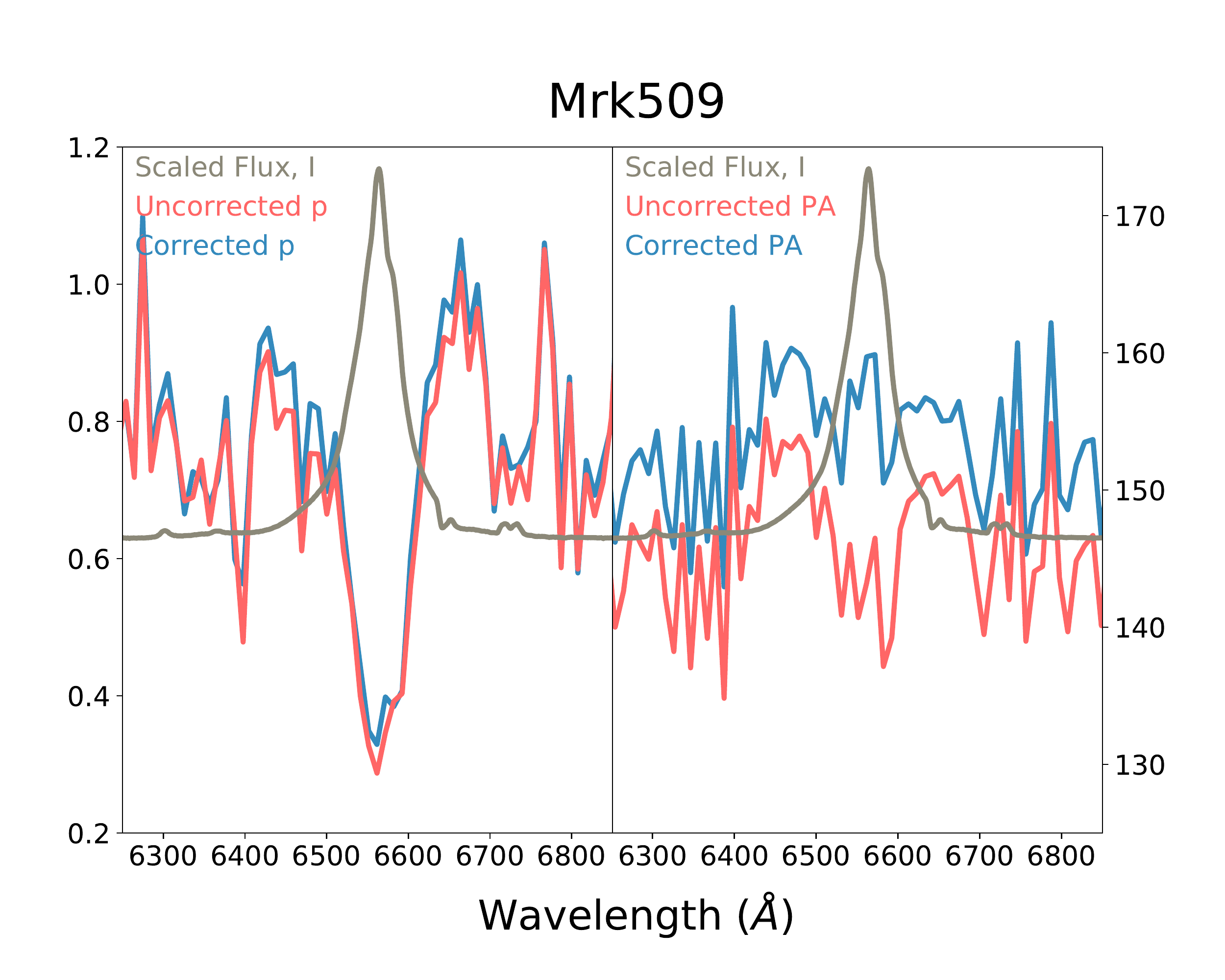}
 \caption{ISP corrected and uncorrected $p$ (left) and PA (right)
   spectra of Mrk\,509 in the H$\alpha$ spectral region.}
 \end{figure} 

For Mrk\,509 the H$\alpha$ and H$\beta$ spectral regions were observed
during the same observing run and therefore can be examined together.
The red spectrum is however of low quality, and we present heavily
binned data in Figures 2 and 4.

Variability in the scattering properties of Mrk\,509 indicates that at
least one scattering component corresponds to a compact region,
probably located close to the BLR. In fact, Young et al.\ (1999)
showed that the polarization level and position angle in the 4350-7100
\AA\ range present a clear correlation in data compiled from 1985 to
1997, with $p$ increasing from $\sim 0.5\%$ at $\sim 140\degr$ to
$\sim 0.9\%$ at $\sim 155\degr$, previous to any correction to the
polarized signal due to an ISP component. Our uncorrected
spectropolarimetric observations for Mrk\,509 with $p \sim 0.9 \pm 0.1
\%$ at $\sim 145 \pm 1\degr$ in the 5300-5600 \AA\ range, are in
agreement with the previous trend. The ISP corrected values are $p =
0.86 \pm 0.04 \%$ and PA $= 150 \pm 1\degr$.

As with NGC\,3783, the $p$ spectrum of Mrk\,509 is characterized by
strong depolarization at the position of the emission lines. In the
$p\times I$ spectrum the FWHM for H$\alpha$ is $\sim 120$ \AA, almost
twice the width of the line seen in total flux, as expected for
equatorial scattering. Contrary to what is observed in NGC\,3783, no
strong break is seen in the continuum at wavelengths $\la 4000$
\AA\ in $p\times I$ flux shown in Figure 2 and 4, but some flattening
is present.

The profiles of the Balmer lines change significantly along the PA
spectrum, starting with a slightly asymmetric bump for H$\alpha$ (the
quality of the data does not allow further characterization), to the
appearance of a trough at the center of the H$\beta$ profile, with
deeper troughs appearing in higher Balmer transitions. Notice that
these troughs cannot be caused by the narrow line components, as it
was already established that the polarization level of the NLR is very
low. The PA shape for the higher-transition Balmer lines resembles the
{\it `$M$-shaped'\/} profile observed in NGC\,3783. This is
particularly clear in the H$\gamma$, H$\delta$ and H$\epsilon$
transitions. The pattern is, however, more symmetric than in the case
of NGC\,3783, with the central trough and the shoulders of the $M$
feature showing similar amplitudes. 

Several previous spectropolarimetric observations of Mrk\,509 have
been published. In particular, Smith et al.\ (2002) presented three
low signal-to-noise observations obtained during 1996 and 1997, Schmid
et al.\ (2000) presented high signal-to-noise data of the H$\alpha$
region obtained in 1999, and Afanasiev et al.~(2019) presented more
recent H$\alpha$ observations obtained in 2014. Hence, Smith's are the
earliest, followed by Schmid and our observations (taken in 2006),
while Afanasiev's correspond to observations obtained about 8 years
later than ours. Comparing this sequence of data shows large
variations in the polarimetric properties of the source, particularly
in the $p$ spectrum. Since neither Smith et al.\ (2002), Schmid et
al.\ (2000) nor Afanasiev et al.~(2019) corrected their observations
for ISP effects, it is important first to look at the changes
introduced by this correction. As can be seen in Figures 1 and 2, and
in more detail in Figure 5, while the ISP correction flattens the PA
signal across the H$\alpha$ emission line to a small degree, little
change is seen in $p$. It is therefore safe to compare observations of
this source despite the different treatments of the ISP correction.

Comparing Smith et al.\ (2002) data to that of Schmid et al.\ (2000)
shows a consistent picture, with a strong trough seen in the $p$
spectrum. The PA spectra are also consistent, with clear
depolarization at the center of the H$\alpha$ line, but the exact
morphology of the high velocity wings is very difficult to asses in
Smith's data. Our observations are also consistent with these two
works, although the relative strength of the shoulders seem to be
reversed in our $p$ spectrum when compared with those of Smith's and
Schmid's.

The morphology of the H$\alpha$ PA line can be easily inspected in
Schmid's work and it seems consistent with the equatorial scattering
predicted profile, although the same can be said of our PA profile
{\em before\/} ISP correction, as shown in Figure 5. The lower quality
of our data makes the comparison hard to asses more thoroughly,
however. As we have already pointed out, it is in the higher Balmer
transitions where the departure of the PA profiles from the simplest
form of equatorial scattering becomes increasingly clear.

Things appear different in Afanasiev's data. The source seems to be
consistent with a pure equatorial scattering, Smith-type profile in
the $p$ profile, with very little evidence of a redward shoulder which
should be found at $\sim 6675$ \AA\ or $\sim +300$ km s$^{-1}$. This
is a strong departure from the observations presented by Smith, Schmid
and our work. It can be postulated, therefore, that the episode of
activity that caused the $M$-type profiles in Mrk\,509 was not longer
present by 2014.

\section{Modelling}

Recently, Lira et al.~(2020) have explored the BLR polarised signal
for different geometrical and dynamical scenarios, and have found that
equatorial scattering can give rise to a wide range of PA profiles. In
particular, they have been able to obtain profiles similar to those
first proposed by Smith et al.~(2002, 2004), but also $M$-type
profiles that resemble those observed in NGC\,3783 and Mrk\,509. These
last models required that 1) the BLR and a scattering media are
coincident, 2) both regions undergo Keplerian rotation, 3) the
scatterer must be optically thin in the polar direction but offer
enough optical depth to photons escaping at low angles with respect to
the disk geometry, and 4) the scattering medium presents a significant
outflowing velocity.

\begin{figure*}
\centering
\includegraphics[scale=0.7,trim=0 0 0 0]{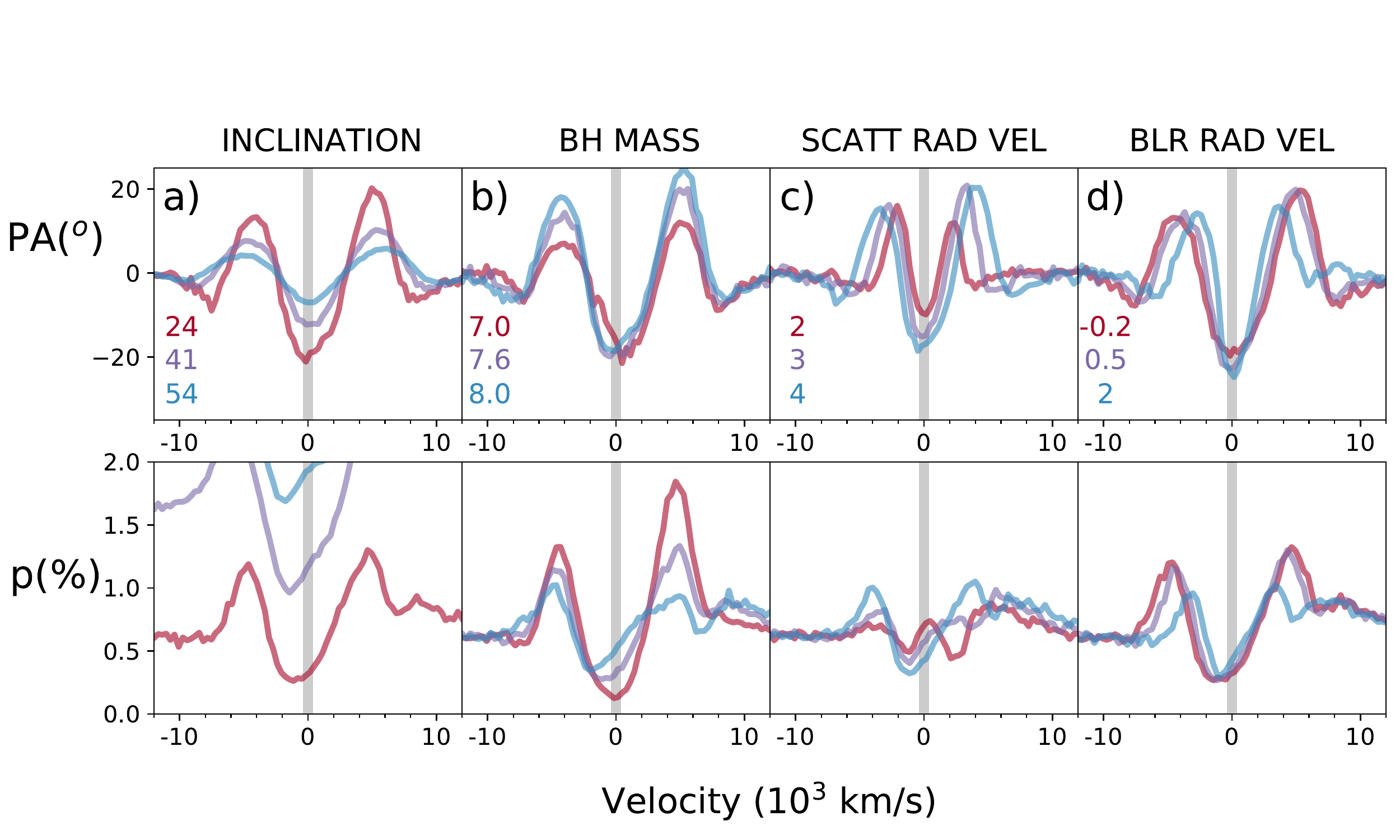}
\includegraphics[scale=0.7,trim=0 0 0 0]{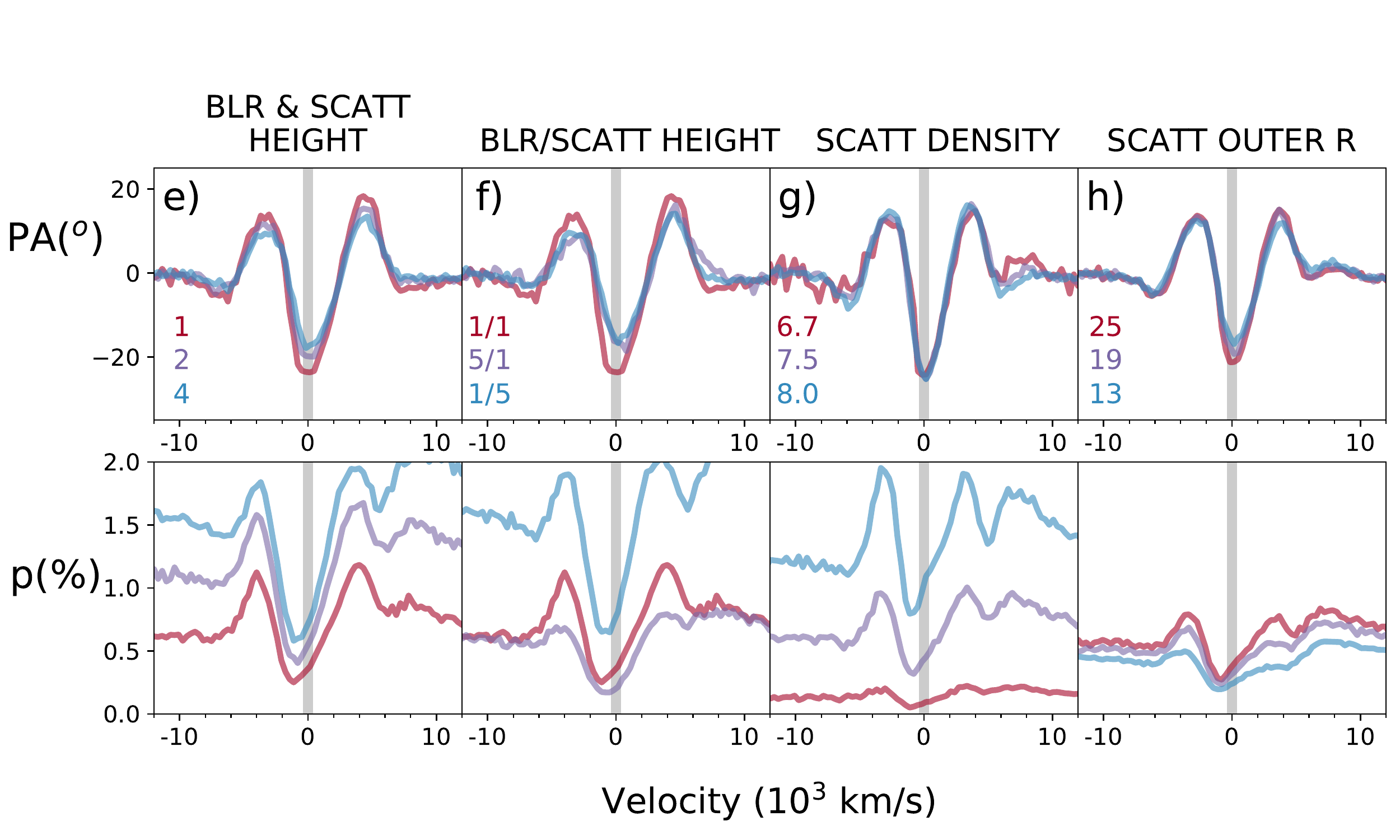}
\caption{Changes in PA and $p$ spectra as a function of the following model parameters:
  {\bf a)} inclination (viewing) angle (in degrees; notice that the $p$ spectrum for the 54 degree inclination is off the plot),
  {\bf b)} log of BH mass (in $M_{\odot}$),
  {\bf c)} scatterer radial velocity (in $10^3$ km/s),
  {\bf d)} BLR radial velocity (in $10^3$ km/s),
  {\bf e)} BLR and scatterer height (in milli-pc),
  {\bf e)} BLR/scatterer height ratio (in milli-pc),
  {\bf g)} log of scatterer density (in cm$^{-3}$),
  {\bf h)} scatterer outer radius (in milli-pc).
  Color-coded values are presented in each panel. For further details see Table 1.}
\end{figure*}

\subsection{Exploration of Parameter Space}

\begin{table*}
\caption{Parameter Values for Figure 6}
\begin{tabular}{llllllllllllll} \hline
         &\multicolumn{2}{c}{General Parameters} && \multicolumn{4}{c}{BLR Parameters}                                      && \multicolumn{5}{c}{Scatterer Parameters} \\ \cline{2-3} \cline{5-8} \cline{10-14}
Panel    & Inclination    & log $M_{\rm BH}$   && $R_{\rm in}$& $R_{\rm out}$ & Height    & $v_{\rm rad}$    && $R_{\rm in}$     & $R_{\rm out}$       & Height          & log $\rho_{\rm e}$ & $v_{\rm rad}$ \\
         & degs           & $M_{\odot}$        && mpc        & mpc       & mpc          & $10^3$ km/s     && mpc             & mpc                & mpc             & cm$^{-3}$         & $10^3$ km/s   \\ \hline
{\bf a)} & {\bf 24 41 54} & 7.6               && 4          & 30        & 1            & 0               && 4               & 30                 & 1               & 7.5               & 5            \\ 
{\bf b)} & 24             & {\bf 7.0 7.6 8.0} && 4          & 30        & 1            & 0               && 4               & 30                 & 1               & 7.5               & 5            \\ 
{\bf c)} & 24             & 7.6               && 4          & 30        & 1            & 0               && 4               & 30                 & 1               & 7.5               & {\bf 2 3 4}  \\ 
{\bf d)} & 24             & 7.6               && 4          & 30        & 1            & {\bf -0.2 0.5 2}&& 4               & 30                 & 1               & 7.5               & 5            \\ 
{\bf e)} & 24             & 7.6               && 4          & 30        & {\bf 1 2 4}  & 2               && 4               & 30                 & {\bf 1 2 4}     & 7.5               & 5            \\ 
{\bf f)} & 24             & 7.6               && 4          & 30        & {\bf 1 5 1}  & 2               && 4               & 30                 & {\bf 1 1 5}     & 7.5               & 5            \\ 
{\bf g)} & 24             & 7.6               && 4          & 30        & 1            & 2               && 4               & 30                 & 1               & {\bf 6.7 7.5 8.0} & 5            \\ 
{\bf h)} & 24             & 7.6               && 4          & 30        & 1            & 2               && 4               & {\bf 25 19 13}     & 1               & 7.5               & 5            \\ 
\hline
\end{tabular}
\end{table*}

In this section we will present variations to the modelling carried
out in Lira et al.~(2020) in order to fit our observations.
Essentially, we would like to determine the physical characteristics
of the BLR and the scatterer invoking a model that is able to
reproduce most of the key aspects of the data, such as the behavior
observed in the $p$ and PA spectra at the location of the BLR emission
lines. All models have been obtained using the STOKES software
(Goosmann \& Gaskell 2007, Marin et al.~2012). Other examples of
STOKES modelling and capabilities can be found in Marin et al.~(2015),
Marin (2018), Rojas Lobos et al.~(2018), and Savi{\'c} et al.~(2018,
2020).

One caveat that needs to be explained at this point is that while the
BLR is an optically thick emitting source, STOKES does not allow for
such structures. Instead, once an emitting source is defined, all its
volume is responsible for emitting line photons at random directions.
Hence, the anisotropic nature of the emission from BLR clouds (e.g.,
Pancoast et al., 2014) cannot be incorporated in our models.
Furthermore, as the scattering region is required to be optically thin
in the polar direction, ultra thin BLR and scattering disks are
adopted. This is not a good physical representation of the sources,
but it is required by the current limitations of the modelling. A
better motivated representation would correspond to optically thick
emitting clouds surrounded by a thin atmosphere of outflowing gas.

In Figure 6 we present the PA and $p$ spectra for several variations
of the base model already presented in Lira et al.~(2020). The base
model consists of what is known for NGC\,3783 in terms of its central
black hole and the location of the BLR: a $4 \times 10^7$ M$_{\sun}$
in mass surrounded by a Keplerian-rotating, thin disk-like BLR.  The
line emission generated by this BLR is centered at 6563 \AA\ and has
an intrinsic width of 50 \AA\ or 2286 km/s. The scattering region is
coincident with the BLR, has a density of at least $5 \times 10^6$
electrons per cm$^{-3}$, and presents Keplerian rotation as well as an
equatorial outflowing wind. Crucially, the PA profile from this base
model shows a $M$-type PA profile, characterized by the peaks and
central trough having similar amplitudes (see Lira et al., 2020),
which is close to what is seen in the higher order Balmer lines in
Mrk\,509 but a bad match for the profile seen in NGC\,3783, which is
characterized by a very deep central trough, and so variations to this
base model need to be explored.

In Figure 6, panel {\bf a}, we show that the amplitude of the $M$
feature is a strong function of the line-of-sight angle ($i$), as the
PA is subject to a scaling by a $\cos(i)$ factor ($i \sim 0$
corresponds to a face-on disk). The level of polarization, $p$,
suffers the opposite effect, as a face-on orientation gives a null $p$
value due to complete geometrical cancellation for a perfectly
circular geometry (notice that in panel {\bf a} the $p$ spectrum for a
54 degree inclination angle is off the plot). 

BH mass has a great impact on both, the PA and $p$ spectra (panel {\bf
  b} in Figure 6) for a BLR located at the {\em same physical
  radius\/} from the central BH. The PA spectrum develops stronger
shoulders for massive BHs, with the redwards shoulder being more
prominent. The $p$ line shape shows a strong asymmetric `U' shape for
low BH masses, becoming less prominent and skewed towards the blue for
larger BH masses. Notice that the {\em continuum\/} $p$ level remains
constant regardless of the BH mass.

Panels {\bf c} and {\bf d} in Figure 6 show that the shape of the PA
and $p$ profiles strongly respond to the radial speeds of the BLR and
scatterer. In general terms, as the radial velocity $v_{\rm rad}$ of
the scatterer increases (panel {\bf c}) the $M$ profile gets
amplified, becoming broader and with larger peaks and troughs. In all
cases the continuum $p$ level remains constant, while at the line
position complex, although low-amplitude structure appears. The
$p$ profile develops a $M$-type profile for $v_{\rm rad} > 3000$
km/s. Notice also how the trough of the $p$ profile becomes skewed
towards the blue for the largest $v_{\rm rad}$ value.

Strong changes in PA are also seen for different values of the BLR
radial velocity (panel {\bf d} in Figure 6). The profile becomes wide
and less deep for negative (inflow) velocities. Likewise, it becomes
narrower and deeper for larger outward speeds. Similar trends are seen
in the $p$ spectra.

Panels {\bf e-h} in Figure 6 present the dependency on the disk
height, outer radii, and electron density.  As can be seen, these
changes mostly affect the $p$ spectra, with only small changes
introduced to the PA morphology.

The heights of the BLR and scattering media have a large effect on the
observed value of the continuum of the $p$ spectra. For equal height
of both media (panel {\bf e}), the continuum level rises by $\sim 50\%$
each time the height is doubled (from 0.001 to 0.002 to 0.004 pc). The
ratio between the thickness of the BLR and scattering media shows that
while a BLR spatially thinner than the scatterer boosts the $p$
continuum, the opposite happens when the the BLR is thicker than the
scatterer (panel {\bf f}). Interestingly, the profile of the line in
the $p$ spectrum shows reduced shoulders in this last case, which
seems a closer match to the Mrk\,509 data.

It is found that the physical thickness of the disk is somewhat
degenerate with the electron density $\rho_{\rm e}$ (panel {\bf g})
in that both parameters largely control the level of continuum
polarization $p$ without altering the PA spectrum. Changes in electron
density, however, introduces strong changes in the line observed in
the $p$ spectrum, with the amplitude of the profile becoming larger
for higher values of $\rho_{\rm e}$, and also resulting in the
appearance of very strong and sharp shoulders, which is not seen by
increasing the BLR and/or scatterer height. Notice also that any
structure with a height larger than 0.001 pc will become optically
thick in the vertical direction for electron densities above
$\rho_{\rm e} = 10^{8}$ cm$^{-3}$.

The radial location of the scattering medium also impacts the observed
profiles, but to a lesser degree. Panel {\bf h} of Figure 6 presents
models for a scatterer starting at 0.005 pc and with outer edges at
0.025, 0.019 and 0.013 pc. It shows that the impact in the PA profiles
is insignificant, while some changes are observed in the $p$
spectra. As expected, the larger the overlap between BLR and
scatterer, the higher the polarization fraction.

Table 1 summarizes the different parameter values presented in Figure
6. Variations of several parameters, however, do not simply correspond
to the combination of the different features shown in Figure 6 as this
is clearly a non-linear problem. In particular, it was found that
varying velocity radial profiles were needed to properly account for
the observed properties when modelling NGC\,3783 and Mrk\,509 in the
following Sections.

\subsection{Spectropolarimetric modelling of NGC\,3783 and Mrk\,509}

\begin{figure*}
\centering
\includegraphics[scale=1.0,trim=20 0 0 0]{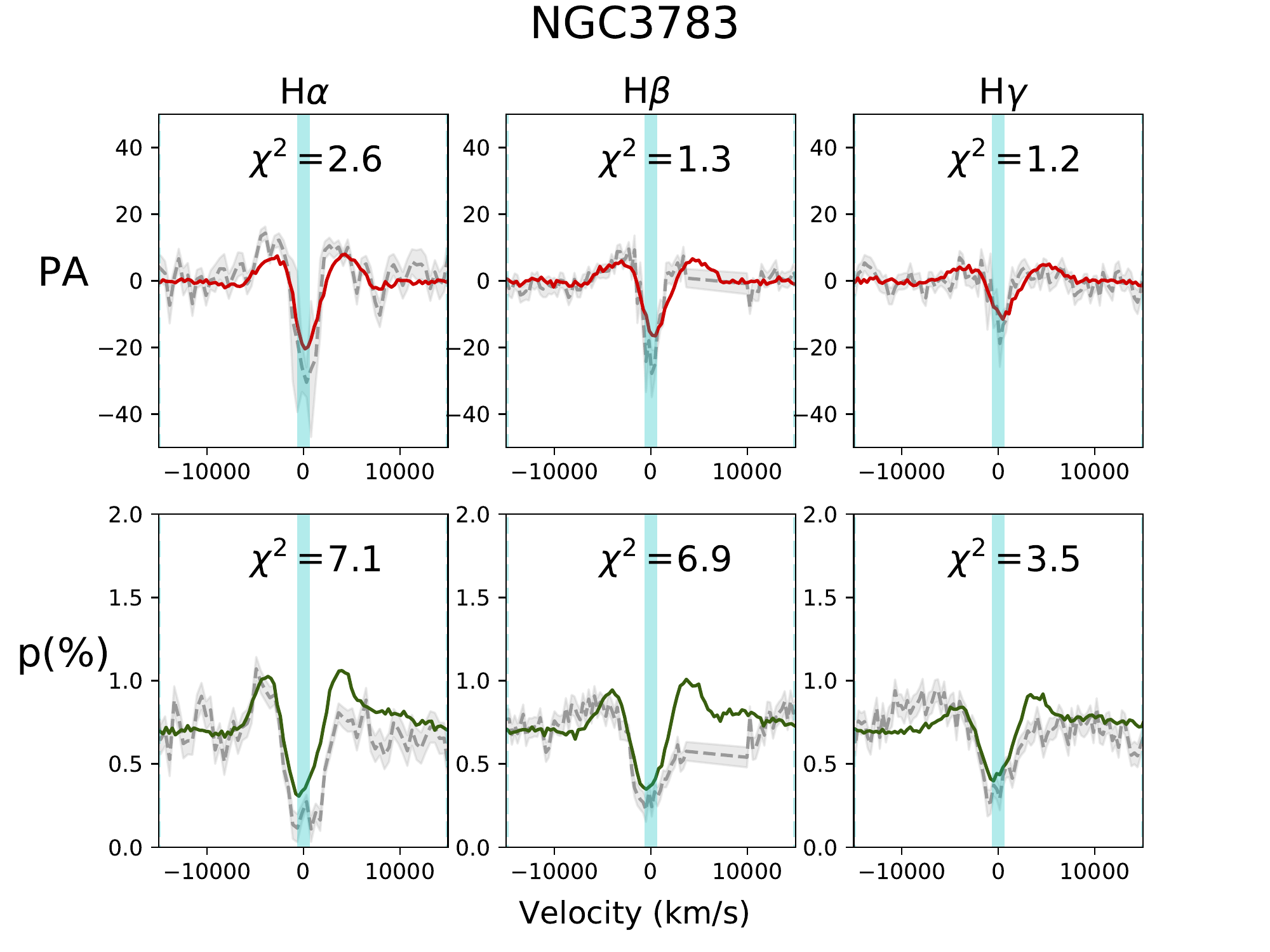}%
\caption{Best fit model for NGC\,3783 PA (top row) and $p$ spectra
  (bottom row). Data are shown with gray dashed lines and errors are
  presented as a gray region around the data. A best fit model
  corresponds to the model with the best sum of $\chi^2$ from all fits
  to H$\alpha$, H$\beta$ and H$\gamma$.  Data between -15000 and
  $+15000$ km/s from the line centers were considered in the fits,
  with the OIII lines masked out bluewards of H$\beta$.}
\end{figure*}

\begin{figure*}
\centering
\includegraphics[scale=1.0,trim=20 0 0 0]{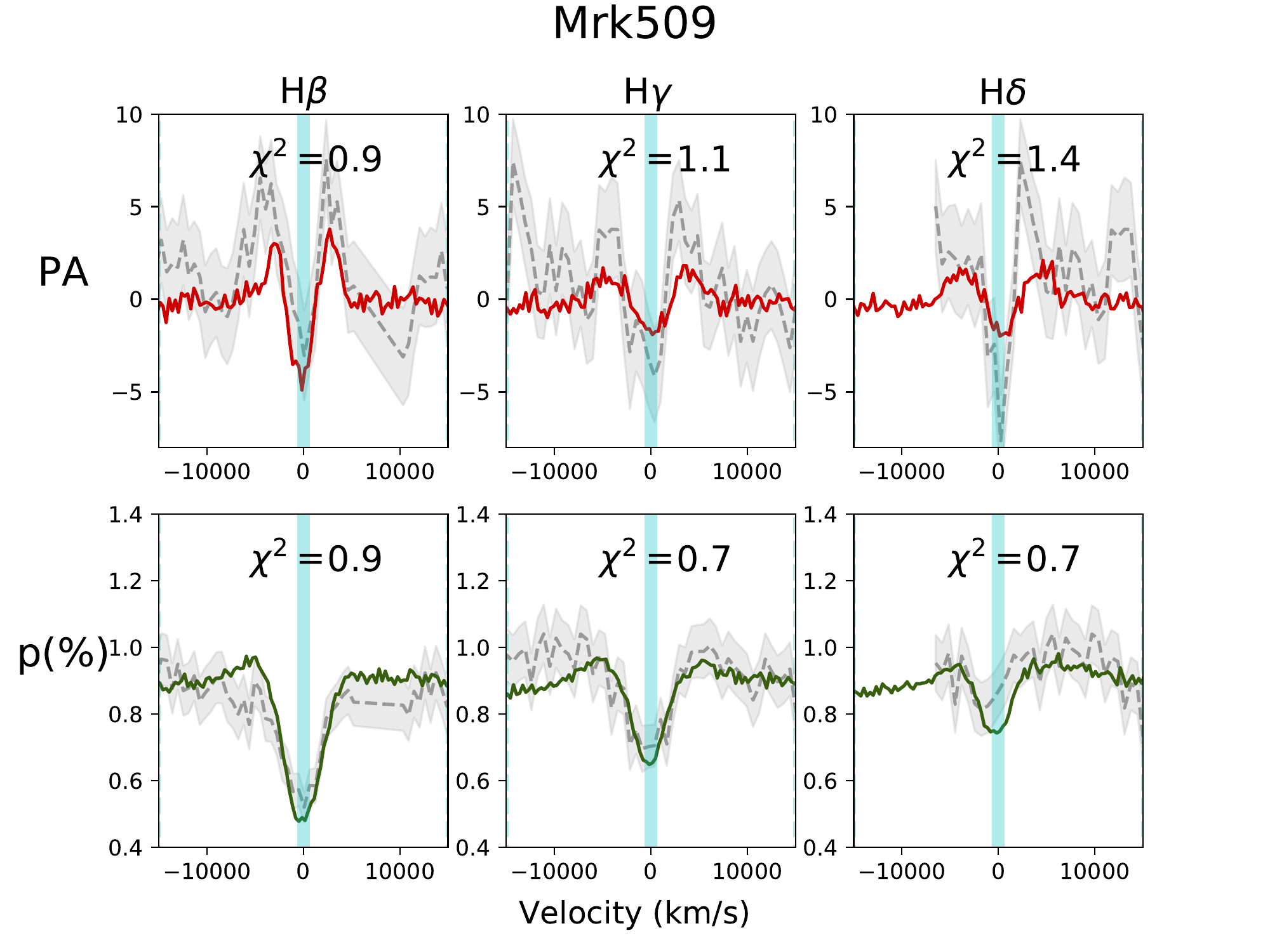}%
\caption{Same as Figure 7 but for Mrk\,509 for the H$\beta$, H$\gamma$
  and H$\delta$ Balmer lines. Notice that the data for this object is
  of lower signal-to-noise than NGC\,3783.}
\end{figure*}

\begin{table*}
\caption{Best Fit Parameters for NGC\,3783 and Mrk\,509}
\begin{tabular}{lll} \hline
Object / Model Parameter Value & Observed Value & Reference \\ \hline
NGC\,3783 & & \\
M$_{\rm BH} = 3.5 \times 10^7$ M$_{\sun}$ & 2.4 - 3.5 $\times 10^7$ M$_{\sun}$ & Onken \& Peterson (2002)\\
Assumed inclination angle = 24 degrees & & \\
BLR innermost radius R$_{\rm in} = 0.003$ pc ($1\times 10^3 R_{\rm Sch}$) & & \\
BLR line emissivity profile $\sim r^{-1/2}$ & & \\
BLR peak emissivity radii (H$\gamma$, H$\beta$, H$\alpha$) = $4, 6, 10 \times 10^{-3}$ pc (1, 2, 3 $\times 10^6\ R_{\rm Sch}$) & H$\beta$ RM results = 6-8$\times 10^{-3}$ pc & Onken \& Peterson (2002)\\
BLR radial velocity = 2500 km/s & & \\
BLR and Scatterer height = 0.001 pc (335 $R_{\rm Sch}$) & & \\
Scatterer innermost radius R$_{\rm in}$ = 0.001 pc ($1\times 10^3 R_{\rm Sch}$) & & \\
Scatterer electron density n$_{\rm e}$ = $3 \times 10^7$ cm$^{-3}$ & & \\
Scatterer vertical Thompson opacity $\tau = 0.07$ & & \\
Mean free path of photons = 0.015 pc ($5\times 10^3 R_{\rm Sch}$) & & \\
Scatterer outflow velocity = 8000 km/s at 0.002 pc; 4000 km/s at 0.035 pc & & \\
\hline
Mrk\,509 & & \\
M$_{\rm BH} = 1.4 \times 10^8$ M$_{\sun}$ & $1.4 \times 10^8$ M$_{\sun}$ & Peterson et al.~(2004)\\
Assumed inclination angle = 24 degrees & & \\
BLR innermost radius R$_{\rm in} = 0.02$ pc ($2\times 10^3 R_{\rm Sch}$ ) & & \\
BLR line emissivity profile $\sim r^{-1/2}$ & & \\
BLR peak emissivity radii (H$\delta$, H$\gamma$, H$\beta$) = $2, 6, 10 \times 10^{-2}$ pc (2, 6, 10 $\times 10^5\ R_{\rm Sch}$) & H$\beta$ RM results = 6-8$\times 10^{-2}$ pc & Peterson et al.~(2004)\\
BLR radial velocity =  2000 km/s at 0.02 pc; -1500 km/s at 0.1 pc & & \\
BLR and Scatterer height = 0.0015 and 0.003-0.005 pc (150 and 300-500, $R_{\rm Sch}$) & & \\
Scatterer innermost radius R$_{\rm in}$ = 0.02 pc ($2\times 10^3 R_{\rm Sch}$) & & \\
Scatterer electron density n$_{\rm e}$ = $4 \times 10^7$ cm$^{-3}$ at 0.02 pc; $6 \times 10^6$ cm$^{-3}$ at 0.1 pc & & \\
Scatterer vertical Thompson opacity (at 0.1 pc) $\tau = 0.05$ & & \\
Mean free path of photons (at 0.1 pc) = 0.077 pc ($8\times 10^3 R_{\rm Sch}$) & & \\
Scatterer outflow velocity = 6000 km/s at 0.02 pc; 1000 km/s at 0.15 pc & & \\
\hline
\end{tabular}
\end{table*}

In what follows we will present the best fit modelling obtained for
several emission lines observed in polarimetric mode for NGC\,3783 and
Mrk\,509. We will not attempt to give a physical interpretation of how
the polarized signal arises as a result of the different adopted
parameters, as Lira et al.~(2020) showed that the complex nature of
these models makes a straightforward analysis a rather difficult
task. Instead, we will interprete the best fit parameters in the
context of what is already known about these two sources.

The parameter space was explored carefully in order to determine the
``best fit'' models. Unfortunately, a MCMC procedure was not adopted
because of time restrictions given the large number of possible
variables and the expense of every single simulation. Instead, we
manually tested different parameter combinations after identifying how
the models responded to parameter changes, as already shown in Figure
6. The most important observational constraints come from the
Reverberation Mapping (RM) results for both sources, as presented by
Onken \& Peterson et al.~(2002) for NGC\,3783 and Peterson et
al.~(2004) for Mrk\,509.

One aspect of the modelling that is important to address now is that
different emissivity profiles as a function of radius will be adopted
for the different lines in the Balmer series. This seems
counterintuitive at first, since all of them are produced by electron
recombination to the same ion, H$^{+}$. However, it is well
established that for very high electron and flux densities, as is the
case of the BLR, the Balmer lines are not produced under the same
physical conditions. This is because the populations of excited states
of neutral Hydrogen not only cannot be neglected, but also are
proportional to the flux of ionizing photons (e.g., Ferland, Netzer \&
Shields 1979; Ferland et al.~1992). An important consequence is that
different Balmer lines are formed at different distances from the
central nucleus, which is clearly demonstrated by reverberation
mapping results, with the weighted mean delays for the first three
lines in the series found to be
$\tau$(H$\alpha$):$\tau$(H$\beta$):$\tau$(H$\gamma$) $=
1.54:1.00:0.61$ (Bentz et al.~2010). It is then natural that different
emissivity radial profiles are also adopted when analyzing different
lines of the Balmer series.

The goodness of the fits were checked applying $\chi^2$ tests to the
features observed in the normalized Stokes Q/I ($q$) and U/I ($u$)
data. Using $q$ and $u$ instead of the PA and $p$ is less problematic
since it does not suffer from degeneracies, $p$ been always positive
(while $q$ and $u$ can be positive or negative) and PA also presenting
a sign degeneracy. By normalizing the Q and U spectra by the total
intensity I, the results become independent of the brightness of the
source (and model, as the strength of the line is given by the number
of simulated photons, which is completely arbitrary) and allows to
examine only the level of fractional polarized flux. In what follows
we present and discuss the best fit results as seen in the $p$ and PA
spectra, but the reader can also find the $q$ and $u$ fitting results
in the Appendix.

Fits to the data were obtained within $\pm$ 15000 km/s of the line
centers, which ensures that the line features as well as the continuum
level were properly taken into account. The intrinsic PA value
corresponding to the continuum level was subtracted from the PA
spectra so that it became consistent with a position of 0 degrees. The
regions corresponding to the OIII emission lines near H$\beta$ have
been masked from the fitting procedure for both sources. Physical
parameters of the BLR and scattering regions are found in Table 2. As
was already discussed in the previous section, the outer radii of the
scatterer region does not change the modelling results significantly
as long as the value is at least a few times that of the inner radii
and hence are not included in Table 2.

One observational constraint for both, NGC\,3783 and Mrk\,509 is that
the level of fractional polarization $p$ remains relatively constant
throughout most of the observed wavelength range (although it shows a
slight increase towards the blue in Mrk\,509), except below
4000\AA\ where it presents a clear drop, as previously discussed. This
means that the height and electron density of the BLR and scattering
region cannot vary strongly with radii, otherwise the level of the
continuum $p$ would not stay at the same level, unless opposite
effects just cancel each other.

\subsection{Best fit to NGC\,3783}

The best fit solutions for the H$\alpha$, H$\beta$ and H$\gamma$ PA
and $p$ spectra are presented in Figure 7. $\chi^2$ values are given
for each Balmer line and spectrum. The only difference between the
modelling of the H$\alpha$, H$\beta$ and H$\gamma$ emission lines is
in their emissivity profiles. They are all characterized by a $\sim
r^{-1/2}$ function but have different starting radii and
normalizations. All other parameters defining the BLR and scatterer
remained fixed for the best fit model shown in Figure 7.

NGC\,3783 shows a persistent decrease in the amplitude of the $M$-type
profiles towards higher transition Balmer lines, as seen in the PA as
well as the $p$ spectra. The shoulders of the $M$ features also become
less sharp. On the other hand, the $p$ continuum remains largely
constant. The main challenge was to find a parameter combination that
would yield sufficient $p$ continuum without too sharp and large
shoulders in the PA and $p$ lines but still a large trough for the PA
$M$ profile.

It is found that a deep trough in the line profiles can be achieved by
the combination of a small BH mass and large radial outflows in both,
the BLR and scattering medium. The scatterer wind velocity changes
from 8000 km/s close to the nucleus to 4000 km/s at 0.03 pc in
radius. This profile was modeled as an equally spaced step function
with steps of 1000 km/s. The BLR wind velocity was explored in steps
of 500 km/s and it did not vary with radius. Gaussian and exponential
emissivity profiles were also tested besides the $\sim r^{-1/2}$ law,
but gave poor results. The width of the scattering region is found to
be 0.4 times its innermost radius, although we remind the reader that
this width is more representative of the width of an atmosphere around
the BLR region than an actual disc width. The innermost radius of the
scattering region was found to be 0.003 pc, half the distance where
the accretion disc B-band emission is seen to peak, and 2.5 times
smaller than the outskirts of the disc as detected in the J-band (Lira
et al., 2011), which is also coincident with the RM lag for H$\beta$
(Onken \& Peterson 2002). Interestingly, using polarization RM for
NGC\,4151 it has been shown that the location of the equatorial
scatterer is found to coincide with the lag measurements for CIV and
H$\beta$ (Gaskell et al.~2012).

In general, the models better reproduce the PA signatures than the
structure observed in the $p$ spectra, in particular the continuum $p$
in the vicinity of the line. This could be due to the limiting
geometrical options available to our modelling, as already discussed
above. Besides, no flaring or warping of the Scatterer can be
explored. Still, the model follows well the general shape of the
troughs and it shows the slight asymmetry observed at the center of
the line, particularly in the profiles of H$\beta$ and H$\gamma$.

\subsection{Best fit to Mrk\,509}

The PA and $p$ line features in Mrk\,509 are broader and show less
amplitude than those seen in NGC\,3783. As already discussed in
section 5.1, it would seem that the BH mass would be sufficient to
explain the differences in PA profile between NGC\,3783 and
Mrk\,509. However, just as their BH masses differ by a factor of $\sim
5$, the location of their BLRs differ by a factor of $\sim 10$ (Onken
\& Peterson 2002, Peterson et al.~2004). Normalizing by their
Schwarzschild radii, the BLRs are found at $2 \times 10^3$ and $7
\times 10^3\ R_{\rm Sch}$ for NGC\,3783 and Mrk\,509, respectively. As
a result, the Keplerian velocities imparted by the BHs at the distance
of the BLR formation are rather similar in both objects.

Mrk\,509 does not show an $M$-type PA profile in H$\alpha$ (see
Figures 2, 4 and 5 and the discussion in Section 3.3). Instead, the
shape seems to be somewhat closer to the profile proposed by Smith et
al.~(2005). In H$\beta$ a clear $M$ PA profile is present and the
troughs become more and more prominent in higher order Balmer
lines. Given the low signal-to-noise of the H$\alpha$, in Mrk\,509 we
will model the H$\beta$, H$\gamma$ and H$\delta$ observed polarized
features.

The observations show that while the PA profile seems to show an
increase in trough amplitude when moving from H$\beta$ to higher order
Balmer transitions (H$\gamma$, H$\delta$ and further), the opposite
trend is seen in the $p$ line profile. Unfortunately, we were not able
to find a unique BLR--scatterer model that would reproduce this
behaviour. The final model presented in Figure 8 shows a successful
fit the the features observed in the $p$ spectra (continuum level and
line features) but which is not able to follow the PA profile changes
seen across Balmer lines.

Comparison between our models presented in section 5.1 and the PA
profiles in Mrk\,509 would suggest that the BLR radial velocity should
change from an outflow at the radius where higher order Balmer lines
are produced to an inflow at the distance where H$\alpha$ arises (see
Figure 6, panel {\bf d}). This however, does not take into account
that photons are not scattered {\em in-situ\/}, but where the optical
depth in the scattering medium becomes $\sim 1$. Choosing different
scatterer electron densities the location where this condition is met
can be modified. To obtain the correct $p$ line profile and $p$
continuum level, we found that the required parameters are: a
$\rho_{\rm e} \sim$ a few times $10^6$ cm$^{-3}$ and a disk height of
at least 0.005 pc. Therefore the mean free path of photons along the
mid-plane of the scattering disk is $\sim 0.1$ pc and changes in the
velocity field would be smeared within this spatial scale. Since RM
results show that all Balmer emission lines should be produced within
$\sim 0.1$ pc from the nucleus in Mrk\,509, for $\rho_{\rm e} \sim
10^6$ cm$^{-3}$ photons will travel freely from regions characterized
by inflowing motions to those characterized by outflowing motions
without preserving a characteristic kinematics.

In order to decrease the level of smearing of the distinct kinematics
within the BLR and scattering disk, we opted for models where larger
electron densities were found at smaller radii. Hence those regions
became more optically thick and photons were scattered closer to the
characteristic radius where they were produced. This strategy reduced
by about half the characteristic radii inside which higher order
Balmer lines were scattered, but it was not effective enough to solve
our problem of the mismatch between the behaviour of the PA and $p$
spectra. MCMC exploration of the parameter space might solve this
issue but it is beyond what can be presently done.

The final model is presented in Figure 8. As with NGC\,3783,
emissivity profiles were described by a $r^{-1/2}$ law with different
starting radii for different Balmer lines. To isolate regions with
different kinematics, the starting (and peak) emissivity for H$\beta$
was chosen to be 0.1 pc, slightly larger than the 0.06-0.08 pc
determined from RM results. The scatterer density decreases from $4
\times 10^{7}$ cm$^{-3}$ at 0.02 pc to $6 \times 10^{6}$ cm$^{-3}$ at
0.1 pc and it stays constant at larger radii. The BLR radial velocity
also varies from an outflow of 2000 km/s in the inner part to an
inflow of -1500 km/s at 0.1 pc. The scatterer, on the other hand,
always presents an outflowing radial motion with speeds of 6000 km/s
in the inner part and reaching a velocity of 1000 km/s at $\sim$ 0.15
pc. Keeping the scatterer and BLR compact ($< 0.15$ pc) resulted in
less pronounced shoulders around the $p$ profile. The typical width of
the scattering region is found to be 0.2 times its innermost radius,
half the value found for NGC\,3783.

\section{Discussion}

One of the most remarkable results from our spectropolarimetric
observations is the change in the line profiles along the Balmer
sequence, as observed in Mrk\,509. This clearly reinforces the notion
that the BLR is an extended source and that physical and dynamical
conditions can rapidly change with radius in AGN nuclei. Furthermore,
it demonstrates that powerful nuclear winds might exist in many Seyfert
galaxies but their presence might only be revealed by observing high
order Balmer lines in polarized light.

The signatures of inflows and particularly outflows have been found in
many AGN. Their presence is also invoked by numerical simulations that
require the AGN to modulate the star formation rate in their host
galaxies (e.g., Harrison 2017, Veilleux et al.~2020), as well as
holistic models of the AGN phenomenology (Elvis 2000, 2017). The
so-called Warm Absorbers represent one of the most commonly observed
nuclear winds: photoionized gas distances of $\sim 1-1000$ pc,
outflowing with velocities of hundreds of km/s, are detected in the UV
and X-ray spectra of $\sim 50\%$ of Seyfert nuclei (Crenshaw \&
Kraemer 1999). Nuclear winds with more extreme properties, detected
much closer to the accretion disk and with velocities of thousands of
km/s, have been observed in only a few sources such as NGC\,5548
(Kaastra et al.~2014), NGC\,985 (Ebrero et al.~2016), Mrk\,335
(Longinotti et al.~2019) and NGC\,3783 (Mehdipour et al.~2017, Kriss
et al.~2019). They have been interpreted as clumpy outflows associated
to the accretion disk (Laha et al.~2021), however, their exact nature
remains unclear. The scarcity of these results might change with
further spectropolarimetric observations of Seyfert 1 nuclei, while
the larger collective area of future 30m-class telescopes will be able
to probe this in many nearby AGN.

We have also shown that the modelling of the data obtained for
NGC\,3783 and Mrk\,509 offers the opportunity to put significant
constraints into the properties of these two sources. Even though we
have not been able to fully explore the parameter space that governs
the emission from these nuclei, and our models still suffer from many
simplifications, we have shown that nuclear high-velocity winds are
necessary to explain the observations. Our exploration of the
parameter space demonstrate that variations in geometrical and
dynamical quantities that characterize the BLR and the scatterer
leaves a clear imprint in the PA and $p$ spectra. Besides, as already
discussed in Lira et al.\ (2020), our modelling shows that in the
absence of a strong radial wind the profile of the lines seen in
spectropolarimetric observations reverts to that proposed by Smith et
al.\ (2002, 2004, 2005).

As already mentioned, recent Swift, XMM-Newton, NuSTAR and HST
observations of NGC\,3783 have detected the presence of a nuclear wind
in this source from the follow up of an obscuring event taking place
in December 2016 (Mehdipour et al.~2017). The event allowed for the
findings of several new absorption features that revealed the
existence of outflowing material with velocities of a few thousand
km/s at distance of about 10 light days ($\sim 10^{-3}$ pc), in
excellent agreement with the results from our STOKES modelling. The
changing features in the spectropolarimetric observations of Mrk\,509
also demonstrate that these winds might be transient phenomena.
Obscuring events such as the one observed in NGC\,3738 are extremely
rare, and reinforces the idea that spectropolarimetry might be the
best tool to search for powerful nuclear winds in AGN.

\section{Data Availability Statements}

The raw data underlying this article can be dowloaded from the ESO
Archive. The processed data will be shared on reasonable request to the
corresponding author.

\section*{Acknowledgments}

PL acknowledges funding from Fondecyt Project \#1201748.
MK acknowledges support from JSPS grant 20K04029.

\appendix

\section{Best-fit results in $q$ and $u$ space}

In the same manner that the PA spectral continuum level was taken to
cero degrees in order to be compared with model results in Section
4.2, the $q$ and $u$ spectra were {\em rotated\/} so that the
continuum level in the $u$ spectra corresponded to cero flux (e.g.,
Landi Degl'Innocenti, Bagnulo \& Fossati, 2007). These new spectra are
denoted as $q'$ and $u'$.

The $q'$ and $u'$ spectra for NGC\,3783 and Mrk\,509 are are shown in
Figures A1 and A2, respectively. The fractional fluxes are expressed in
percentage. For NGC\,3783 $q'$ and $u'$ look remarkably similar, while
for Mrk\,509 the $q'$ spectrum shows more strongly polarized lines and
more symmetric profiles than the $u'$ spectrum.

The best-fit models to the $q'$ and $u'$ spectra are overploted and
the best-fit $\chi^2$ values for each line are shown in the
subpanels. NGC\,3783 presents rather high values of $\chi^2$ and
several wiggles seen in the data are not seen in the model and
viceversa. This could be due to non-negligible atmospheric
depolarization (Section 2.3), a non-perfect rotation of the $q$ and
$u$ spectra, and shortcomings in our modelling capabilities, among
other. The lower signal-to-noise data for Mrk\,509 results in lower
$\chi^2$ values.

\begin{figure}
\centering
\includegraphics[scale=0.5,trim=40 0 0 0]{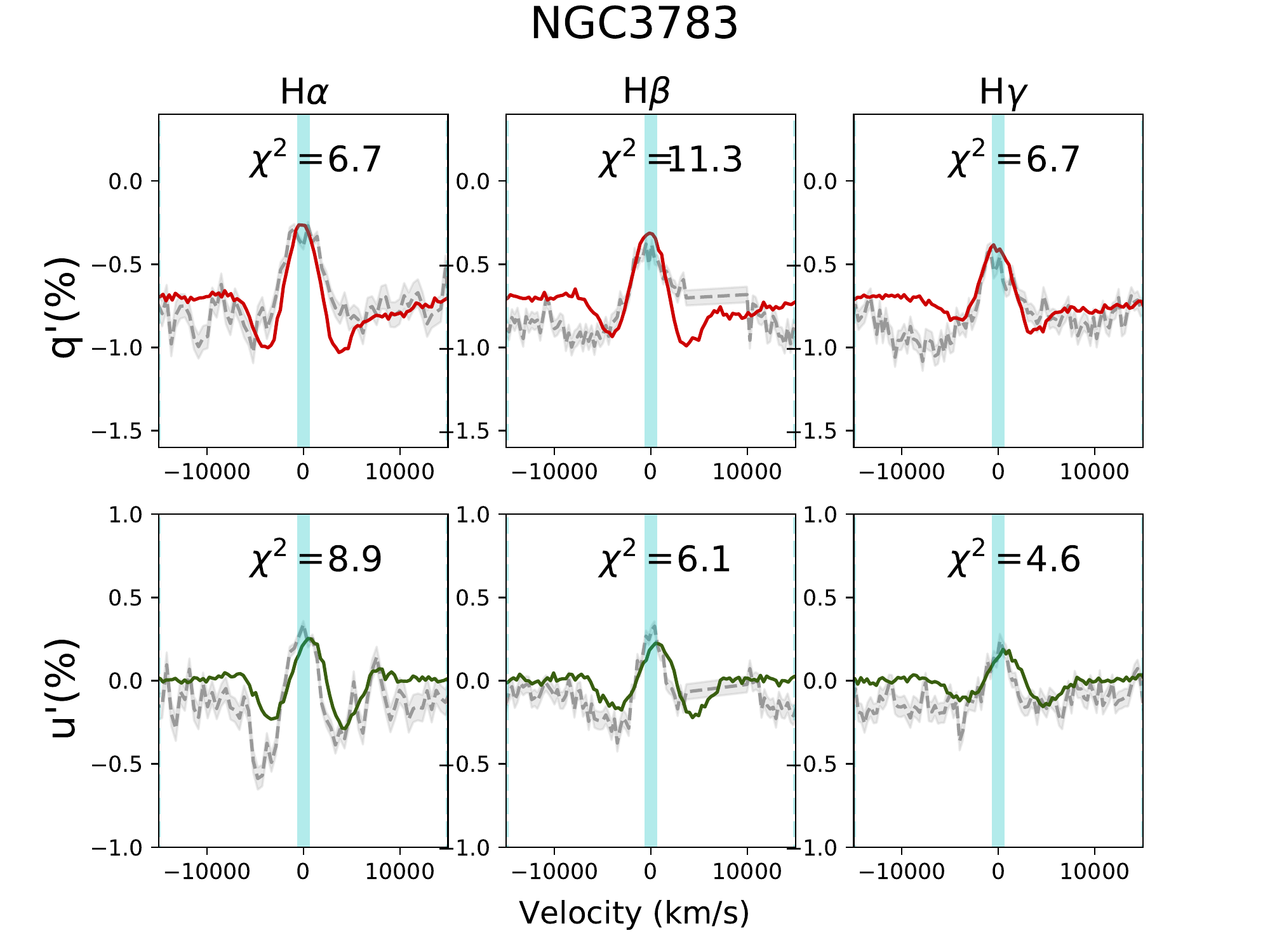}%
\caption{Best fit model for NGC\,3783 $q'$ and $u'$ spectra (top and
  bottom row, respectively). Data are shown with gray dashed lines and
  errors are presented as a gray region around the data. A best fit
  model corresponds to the model with the best sum of $\chi^2$ from
  all fits to H$\alpha$, H$\beta$ and H$\gamma$. Data between -15000
  and $+15000$ km/s from the line centers were considered in the fits,
  with the OIII lines masked out bluewards of H$\beta$.}
\end{figure}

\begin{figure}
\centering
\includegraphics[scale=0.5,trim=40 0 0 0]{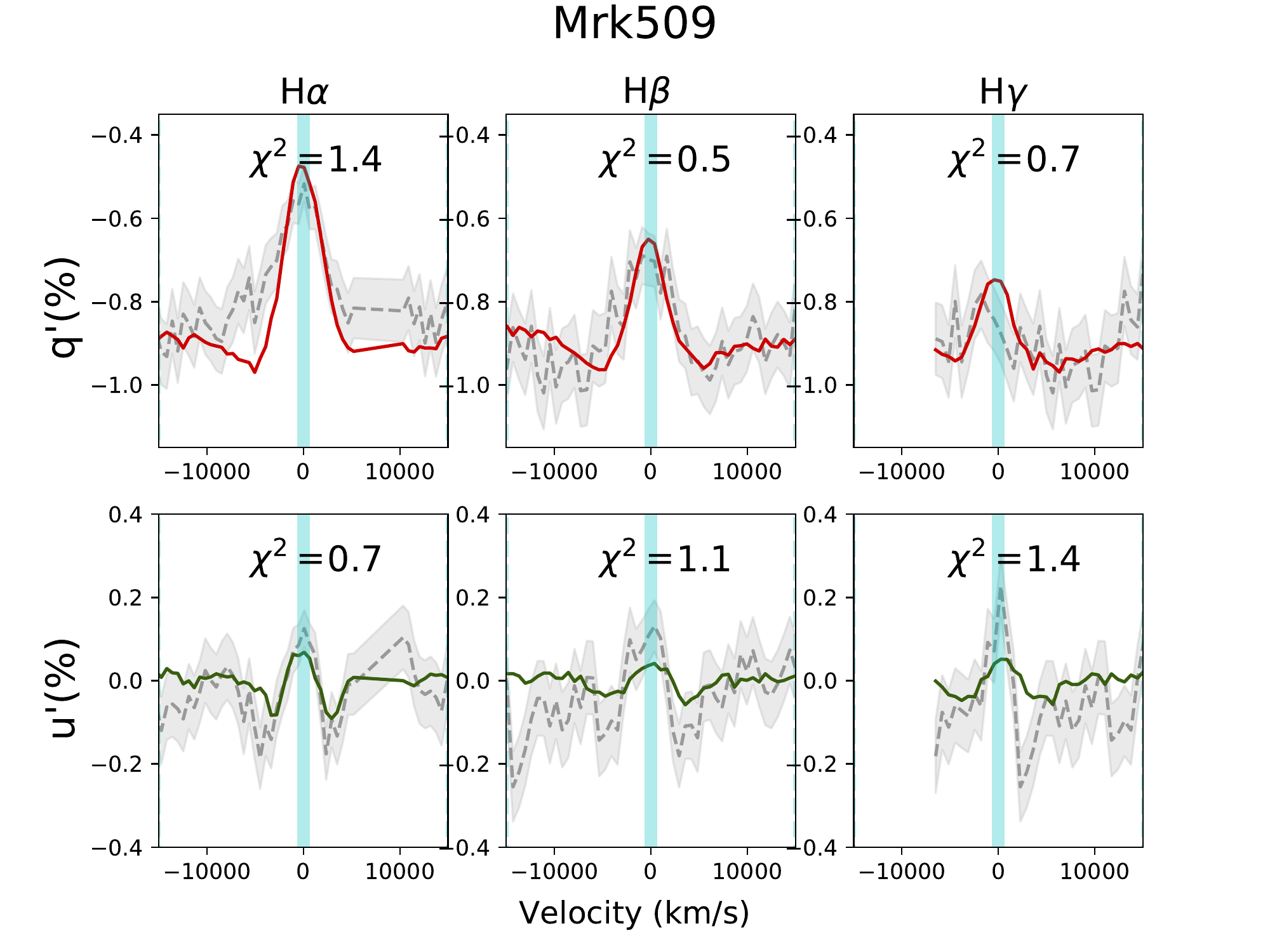}
\caption{Same as Figure A1 but for Mrk\,509 showing the $q'$ and $u'$
  spectra and modelling for the H$\beta$, H$\gamma$ and H$\delta$
  Balmer lines.}
\end{figure}

\bsp

\label{lastpage}

\end{document}